\documentclass[12pt]{article}
\usepackage{amsfonts}
\usepackage{amssymb}
\usepackage{mathrsfs}
\usepackage{graphicx}
\usepackage{amsmath,amsthm,amssymb,amscd}
\usepackage[all]{xy}
\usepackage{subfig}
\usepackage{hyperref}
\usepackage{multirow}
\usepackage{color}
\usepackage{amsfonts}
\usepackage{float}
\usepackage{mathtools,slashed}
\usepackage{bbm}
\usepackage{tikz}
\usepackage{feynmf}

\newcommand{\cf}{{\cal F}}

\newcommand{\nn}{\nonumber}

\def\eqa{\begin{eqnarray}}
\def\eqae{\end{eqnarray}}
\def\eq{\begin{equation}}
\def\eqe{\end{equation}}
\def\be{\begin{equation}}
\def\ee{\end{equation}}
\def\bea{\begin{eqnarray}}
\def\eea{\end{eqnarray}}
\def\ba{\begin{array}}
\def\ea{\end{array}}
\def\bd{\begin{displaymath}}
\def\ed{\end{displaymath}}

\def\tr{{\rm Tr}}
\def\>{\rangle}
\def\<{\langle}

\def\m{\mu}

\def\G{\Gamma}

\def\L{\Lambda}

\numberwithin{equation}{section}

\begin{document}
 \unitlength = 1mm

\begin{titlepage}

\rightline {\small DAMTP-2015-65}
\hfill {\small MCTP-15-18}
\vskip 2cm

\begin{center}
{\Large \bf Dualities in 3D  Large $N$ Vector Models  }
%\vskip .7cm

%{\Large \bf and the Large $N$ Limit }\\
\end{center}

%\vskip .5 cm

\vskip 1 cm
\begin{center}
{  \small\bf Nouman Muteeb${}^{a}$,  Leopoldo A. Pando Zayas${}^{b}$ and Fernando Quevedo${}^{c}$}

\vskip 1 cm

\vspace{.2cm}

{\small \it ${}^{a,b,c}$ The Abdus Salam International Centre for Theoretical Physics, ICTP}\\
{\it Strada Costiera 11, 34014 Trieste, Italy}\\

\vspace{.4cm}
{\small\it ${}^a$ SISSA, Via Bonomea 265, 34136 Trieste, Italy}\\

\vspace{.4cm}
{\small\it ${}^b$  Michigan Center for Theoretical Physics,  Department of Physics}\\
{\it University of Michigan, Ann Arbor, MI 48109, USA}\\

\vspace{.4cm}

{\small\it ${}^c$ DAMTP, CMS, University of Cambridge}\\
{\it Wilberforce Road, Cambridge, CB3 0WA, UK}

\end{center}

\vskip 1.5 cm
\begin{abstract}
Using an explicit path integral approach we derive non-abelian bosonization and duality of 3D systems in the large $N$ limit. We first consider a fermionic  $U(N)$ vector model coupled to level $k$ Chern-Simons theory, following standard techniques we gauge the original global symmetry and impose the corresponding field strength $F_{\mu\nu}$ to vanish introducing a Lagrange multiplier $\Lambda$. Exchanging the order of integrations we obtain the bosonized theory with $\Lambda$ as the propagating field using the large $N$ rather than the previously used large mass limit. Next we follow the same procedure to dualize the scalar $U(N)$ vector  model coupled to Chern-Simons and find its corresponding dual  theory. Finally, we compare the partition functions of the two resulting  theories and find that they agree in the large $N$ limit including a level/rank duality. This provides a constructive evidence for previous proposals on level/rank duality of 3D vector models in the large $N$ limit. We also present a partial analysis at subleading order in large $N$ and find that the duality does not generically hold at this level. 
\end{abstract}

\end{titlepage}

\tableofcontents

%%%%%%%%%%%%%%%%%%%%%%%%%%%%%%%%%%%%%%%%%%%%%%%%%%%%%%%%%%%%%%%
\section{Introduction}
%%%%%%%%%%%%%%%%%%%%%%%%%%%%%%%%%%%%%%%%%%%%%%%%%%%%

One of the most remarkable concepts in theoretical physics  is duality --  the fact that the same physical theory  admits more than one description in terms of  different degrees of freedom. Arguably, one of the most interesting examples of duality is bosonization in two dimensions which evolved from an equivalence of the quantum perturbative expansions \cite{Coleman:1974bu}, to  include the solitonic sector of the theory \cite{Mandelstam:1975hb}, to a more algebraic understanding of the equivalence  in terms of currents \cite{Witten:1983ar}. As a path integral equivalence, two-dimensional bosonization was completely formulated in \cite{Burgess:1993np,Burgess:1994np} wherein through a set of transformations the fermionic partition function was mapped into the bosonic partition function. These serve as one of the very few examples where a duality between two theories can be proved as an equivalence of partition functions using a series of transformations in the path integral rather that making a conjecture and checking some of its implications. 

For many physically important consequences the conjecture and check approach is completely satisfactory as the  AdS/CFT correspondence abundantly shows.  However, conceptually this approach lacks some clarity and finality. Some of these path-integral techniques used to establish two-dimensional bosonization have also been applied to three dimensional systems.  In particular, equivalence, in a certain limit, between very massive fermions and a Chern-Simons theory was established \cite{Burgess:1994tm,Fradkin:1994tt}.

Recently, renewed attention has been paid to three dimensional vector models, that is models with matter fields in the fundamental representation of $U(N)$. This interest is largely motivated by the Higher Spin/Vector Model duality (for a review see \cite{Giombi:2012ms}). In a very precise sense this duality is a baby version of the more established AdS/CFT  correspondence stating the equivalence between strings in AdS${}_5\times S^5$ and ${\cal N}=4$ super Yang Mills in four dimensions. Independently of the original motivation for the renewed interest in 3D vector models, some important properties and relationships among these models have been established  or conjectured using purely field theoretical tools. 

One such example is  a new generalized  level/rank duality for Chern-Simons theories coupled to bosonic or  fermionic matter fields relating  $U(N)$ level $k$ Chern-Simons coupled to critical bosons to $ U(k)$ level $N$ Chern-Simons coupled to free  fermions \cite{Aharony:2011jz,Aharony:2012nh,Aharony:2012ns}. This interesting conjecture was preceded or accompanied by important  work, such as, \cite{Giombi:2011kc,Chang:2012kt,Jain:2012qi,GurAri:2012is}; it has naturally motivated  investigations into various cleanly defined  field theoretic question  for vector models coupled to Chern-Simons theories \cite{Jain:2013gza,Frishman:2013dvg,Bardeen:2014paa,Bardeen:2014qua,Frishman:2014cma,Moshe:2014bja}

In this manuscript we revisit duality transformations in 3D fermionic and bosonic vector models following prescriptions formulated and applied in, for example, \cite{Burgess:1994tm}  using the large $N$ approximation. In section \ref{Sec:Review} we review duality as a bosonization of 3D fermions, we also comment on some extensions and generalizations to more elaborate fermionic theories. In section \ref{Sec:Fermion} we apply the dualization techniques to fermions by gauging a $U(N)$ global  symmetry. Section \ref{Sec:Boson} applies similar techniques to bosonic vector models with scalars in the fundamental representation of $U(N)$.  In section \ref{Sec:Comparison}, we compare the dual actions and verify that they respect the level/rank equivalence of Chern-Simons theories coupled to vector matter fields in the large $N$ limit.  In sections \ref{Sec:BeyondN} and \ref{Sec:EA} we present a discussion of the effective actions beyond the large $N$ limit and find that the duality does not generically hold at this level. We conclude in section \ref{Sec:Conclusions} highlighting a number of interesting open problems that we hope dualization might shed some light on. Some technical considerations are presented in a series of appendices.

%%%%%%%%%%%%%%%%%%%%%%%%%%%%%%%%%%%%%%%%%%%%%%%%%%%%%%%%%%%%%%%%%%%%%%%%%%%%%%%%%%%%%%%%%%%%%%%%%%%%%%%%%%
\section{Review of Duality Transformations for 3d Fermions}\label{Sec:Review}
%%%%%%%%%%%%%%%%%%%%%%%%%%%%%%%%%%%%%%%%%%%%%%%%%%%%%%%%%%%%%%%%%%%%%%%%%%%%%%%%%%%%%%%%%%%%%%%%%%%%%%%%%%%
In this section we review the duality algorithm in the case of 3D fermions \cite{Burgess:1994tm}. Our aim is  to set up the stage for the algorithms and some of the techniques we will apply and extend in this paper. The starting point is the action of a massive fermion coupled to an external field:
\be
\label{Eq:FAbelian}
S_\psi=\int d^3x \bar{\psi}\left(\slashed{\partial}+ m +J_\mu M_\mu\right)\psi,
\ee
where $J_\mu$ is a collection of external fields and $M_\mu$ is a collection of appropriate Dirac matrices, e.g., $J_\mu M^\mu= ia_\mu \gamma^\mu$. The corresponding quantum theory is defined via its Euclidean path integral
\be
Z_{\psi}=\int D\bar{\psi}D\psi \exp(-S_\psi).
\ee

To dualize  the above theory one takes the following steps:

\begin{enumerate}

 \item  Enlarge the fermion theory by gauging the global $U(1)$ symmetry $\psi \to e^{i\theta}\psi$ using a gauge field $A_\mu$.

\item  Constrain the field strength of the corresponding gauge potential, $A_\mu$, to vanish by introducing a Lagrange multiplier, $\Lambda_\mu$.

\end{enumerate}
This procedure leads to the following extended action
\be
\label{Eq:MasterAbelian}
{\cal L}_G={\cal L}_\psi+ i\bar{\psi} \gamma^\mu \psi A_\mu + \epsilon^{\mu \nu \rho} \partial _{\mu}A_{\nu} \Lambda_{\rho},
\ee
where $\Lambda_\rho$ is the Lagrange multiplier enforcing the vanishing of the field strength. The main statement of the duality transformation is that the extended master action (\ref{Eq:MasterAbelian}) is equivalent to the original one (\ref{Eq:FAbelian})and also leads to the dual action:

\begin{itemize}
\item  The original action is recovered by integrating out the Lagrange multiplier which implies that $A_\mu$ is pure gauge.  Integrating over $A_\mu$ which is gauge equivalent to zero amounts to choosing a gauge  where $A_\mu$ is zero leading to the original action.

\item To obtain the dual theory (the bosonized action) one needs to first integrate out the fields $\psi$ and $A_\mu$. The theory that corresponds to $\Lambda_\mu$ is the dual theory.

\end{itemize}

One subtlety in the above dualization procedure resides in global aspects as there might be flat connections that are not smoothly connected to the trivial connection \cite{Alvarez:1993qi}; this obstruction is characterized by the first fundamental group of spacetime. For example, for applications to duality among field theories at finite temperature the above prescription will be insufficient.  In this manuscript we restrict ourselves to situations where this subtlety can be neglected.

The central object is the dual field $\Lambda_\mu$ whose effective action is defined as

\be
\exp\left(iS_\Lambda(\Lambda, J)\right)=\int D\bar{\psi}D\psi DA_\mu \exp\left(iS_G(\psi, \Lambda, J, A) \right).
\ee

The main result of \cite{Burgess:1994tm} consists on evaluating the above integral fairly explicitly. The limit discussed in \cite{Burgess:1994tm} corresponds to an expansion in the inverse fermion mass, $1/m$. The leading term in the $1/m$ expansion is
\bea
\exp\left(i\Gamma_F(A)\right) &=&\int [d\psi] \exp\left(-i\int d^3x \bar{\psi}\gamma^\mu (\partial_\mu -iA_\mu+ m )\psi\right) \nonumber \\
&=& \exp\left(\frac{i}{2}\,\,k \,\int d^3 x\,\epsilon^{\mu\lambda\nu} A_\mu \partial_\lambda A_\nu +\ldots \right),
\eea
where $k ={\rm sign}(m)/(8\pi)^2$ and the ellipsis stands for terms that vanish in the infinite mass $m$ limit, e. g., $\frac{1}{|m|} F_{\mu\nu}F^{\mu\nu}$. Note that we only write explicitly the lowest dimension operator in the action. The last step is integrating with respect to $A_\mu$. Let us consider $J_\mu=a_\mu$, as done in \cite{Burgess:1994tm} to emphasized that we are referring to an applied electromagnetic field.
The integration over the field $A_\mu$ becomes algebraic and the final results is:

\bea
\exp\left(iS_\Lambda(\Lambda, a)\right)&=&\int [dA_\mu]\exp\left(i\Gamma_F(a+A)+i\int d^3 x \epsilon^{\mu \nu \rho} \partial _{\mu}A_{\nu} \Lambda_{\rho} \right), \nonumber \\
S_\Lambda(\Lambda, a)&=& -\int d^3x \epsilon^{\mu \lambda \nu}\left(\frac{1}{2k}\Lambda_\mu \partial_\lambda \Lambda_\nu + a_\mu \partial_\lambda \Lambda_\nu\right).
\eea
Thus, the dual theory, which can also be interpreted as the bosonized version of the original fermionic theory, turns our to be an Abelian Chern-Simons theory coupled to an external field $a_\mu$. It is, therefore,  established that for distances long compared with its Compton wavelength the fermionic theory is dual/equivalent to a Chern-Simons theory.  Note that as a result of the algebraic integration of the gauge field $A_\mu$ the Chern-Simons level of the effective theory appears as $1/k$ with respect to the result of integrating out the fermions. We will pay particular attention to the Chern-Simons level in more general cases.

%%%%%%%%%%%%%%%%%%%%%%
\subsection{Comments and extensions of fermionic duality in 3D}
%%%%%%%%%%%%%%%%%%%%%%
There are several refinements and extensions of the duality transformation presented above, we recall some which will be useful in subsequent sections.  Fradkin and Schaposnik considered the massive Thirring model, i.e., massive fermions with a four-fermion interaction. Using similar techniques they established a duality, to leading order in $1/m$, between the massive Thirring model and Maxwell-Chern-Simons theory \cite{Fradkin:1994tt}. The main technical ingredient that allows to treat the four fermion interaction is its representation as a Gaussian integration over a vector field. The corresponding identity is a rewriting of the four fermion interaction ($J_\mu =\bar{\psi}\gamma_\mu \psi$) as:

\be
\exp\left(\frac{g^2}{2}\int d^3x J^\mu J_\mu\right)=\int D A_\mu \exp\left(-\int d^3 \left(\frac{1}{2}A^\mu A_\mu + g J^\mu A_\mu\right)\right).
\ee
A similar `un-completing the square' identity was used to treat the duality of the 2d Thirring model in  \cite{Burgess:1993np}.

An extension of these dualities to the case of massless fermions was discussed in \cite{Moreno:2013xya}. The issues of non-locality can not be avoided in this case. Namely, the effective action obtained by integrating out the fermions leads to a series of inverse powers of $(\slashed{\partial})^{-1}$ with no mass to regulated those terms \cite{Moreno:2013xya}. There is no natural way to separate scales and therefore terms like $F_{\mu\nu}(-\square)^{-1/2}F^{\mu\nu}$ are not suppressed.

With the hope of clarifying the relevance of the large mass limit, it is also worth pointing out an important perspective in the integration over massless fermions and the appearance of a Chern-Simons theory just like in the infinite mass case \cite{Redlich:1983kn} \cite{Redlich:1983dv}. To evaluate the effective action (integration over fermionic degrees of freedom) at zero fermion mass one requires a regularization because of ultraviolet divergences. The regularization could be taken to be Pauli-Villars $S_{eff}^{reg}[A,m=0]=S_{eff}[A,m=0]-\lim\limits_{M\to \infty}S_{eff}[A,M]$. The Pauli-Villars regularization respects gauge invariance. In the $M\to \infty$ limit the second term leads to the induced Chern-Simon action. Basically, similarly to the situation with the chiral anomaly in four dimensions, maintaining gauge invariance leads to parity breaking. Thus, the appearance of a Chern-Simons term in the context of integrating out massless fermions seems generic. There are many physical applications where the manipulations we describe are central. Recently, for example, they have been applied in the hope of finding an effective field theory for topological insulators \cite{2013PhRvB..87h5132C}.

The extension of the duality transformation to higher dimensions was also described in \cite{Burgess:1994tm}.
Technically, the path integral can be explicitly performed in the large fermion mass limit and the  obtained bosonic theory is found to be that of a rank $(d-1)$  antisymmetric Kalb-Ramond-type gauge potential. In 3d, as we have shown, the result is the Chern-Simons action;  in dimensions higher than three the action is non-local.

Another important extension of the duality transformation reviewed at the beginning of this section is its generalization to the non-Abelian case. Namely, in \cite{Bralic:1995ip}, following steps analogous to those of \cite{Fradkin:1994tt},  the dual of the massive $SU(N)$ Thirring model was constructed. The result, up to $1/m$ corrections, is a complicated action which becomes level $k=1,\,\, SU(N)$ Chern-Simons theory only in the limit of vanishing four-fermion coupling.

%%%%%%%%%%%%%%%%%%%%%%
\section{Duality in $U(N)$ Fermionic Vector Model}\label{Sec:Fermion}
%%%%%%%%%%%%%%%%%%%%%%
Given the global $U(N)$ symmetry of the fermionic vector model we follow the steps  outlined above to construct a dual theory. Namely, we gauge  the symmetry; impose that the gauge field has vanishing curvature and then eliminate the gauge field and the original fundamental fermion writing an action only for the Lagrange multiplier which is, by definition, the dual action.

The starting point is the action of a single fermion in the fundamental representation of $U(N)$

\be
S=\int d^3 x \bigg(\bar{\psi} \slashed{\partial} \psi+V(\bar{\psi}\psi)\bigg).
\ee
We now gauge the global $U(N)$ symmetry and add a Lagrange multiplier, $\Lambda_\rho^a$, imposing  the vanishing of the field strength, $F_{\mu\nu}^a$. The resulting action takes the following form:

\be
\tilde{S}_G= \int d^3x \bigg(\bar{\psi} (\gamma^\mu \partial_\mu -iA_\mu^a T^a \gamma_\mu)\psi +V(\bar{\psi}\psi) + \epsilon^{\mu\nu\rho}F_{\mu\nu}^a\Lambda_\rho^a\bigg).
\ee

%%%%%%%%%%%%%%%%%%%%%%%%%%%%%%%%%%%%%%%%%%%%%%%%%%%%%%%%%%%%%%%%%%%%%%%%%%%%%%5
%\section*{Dualization of fermionic vector model coupled to Chern-Simons }\label{Sec:FermionCS}

Let us now explain a modification with which we shall work. The Higher Spin/Vector Model  dualities such as those originally proposed in \cite{Klebanov:2002ja,Sezgin:2002rt} focused on the singlet sector of vector models. One natural way to zoom in on the singlet sector is through  coupling to a gauge field whose dynamics is governed by a Chern-Simons term.  Moreover, there is a number of exciting new results in the studies of vector models coupled to Chern-Simons. For example, they have been solved in the large $N$ 't Hooft limit in \cite{Giombi:2011kc} (see also \cite{Aharony:2012nh} and \cite{GurAri:2012is}).

%%%%%%%%%%%%%%%%%%%%%%%%%%%%%%%%%%%%%%%%
Therefore, we now consider a theory of a fundamental fermion interacting with $U(N)$ level $k$ Chern-Simons theory. From the point of view of duality one could think of this situation as having a fermion with global $U(N)$ symmetry that one wishes to dualize. Following the standard recipe one would gauge that symmetry and demand that the gauge field have vanishing curvature; the new ingredient would be that one adds a kinetic term in the form of the Chern-Simons action for this gauge field. The complete action we wish to study is, in either perspective:

\bea
S_G&=&\int d^3x \bigg[ \bar{\psi}\gamma^\mu D_\mu \psi + V(\bar{\psi}\psi)+\epsilon^{\mu\nu\rho} F^a_{\mu\nu}\Lambda^a_{\rho} \nonumber \\
&+&\frac{ik}{4\pi}\epsilon^{\mu\nu\rho}\tr \left(A_\mu  \partial_\nu A_\rho  -\frac{2i}{3}A_\mu  A_\nu A_\rho \right) \bigg].
\eea
If we integrate out the Lagrange multiplier $\Lambda^a_{\mu}$ first we simply have that the gauge field is pure gauge and then we can simply fix the gauge that leaves us with the initial action of fermions with $U(N)$ global symmetry. Modulo topological subtleties which we avoid by considering a space with no boundary where the Chern-Simons theory might become dynamical.

We now consider integrating in a different order, namely we would like to integrate in the path integral the fermions and gauge fields and consider the remaining theory of the Lagrange multiplier as the dual theory to the fermionic vector model.

The master theory  is defined as:

\bea
\label{Eq:FermionsMaster}
Z^F&=& \int DA_\mu^a D\bar{\psi}D\psi D\Lambda_\mu\exp\left(-\int d^3x \bigg[\bar{\psi} (\gamma^\mu \partial_\mu -iA_\mu^a T^a\gamma_\mu )\psi +V(\bar{\psi}\psi) + \epsilon^{\mu\nu\rho}F_{\mu\nu}^a\Lambda_\rho^a\right. \nonumber \\
 && \left. +\frac{ik}{4\pi}\epsilon^{\mu\nu\rho}\tr \left(A_\mu  \partial_\nu A_\rho  -\frac{2i}{3}A_\mu  A_\nu A_\rho \right) \bigg]\right) \nonumber \\
&=&\int D\Lambda_\mu Z^F[\Lambda].
\eea

In what follows we will describe approximations to the above path integral with the hope of developing a clear intuition into the properties of $Z^F[\Lambda]$.

%%%%%%%%%%%%%%%%%%%%%%%%%%%%%%%%%%%%%%%%%%%%%%%%%%%%%%%%%%%%%%%%%%%%%%%%%%%%%%%%%%%%%%%%
\subsection{Large mass limit: an insight into the dual theory}
%%%%%%%%%%%%%%%%%%%%%%%%%%%%%%%%%%%%%%%%%%%%%%%%%%%%%%%%%%%%%%%%%%%%%%%%%%%%%%%%%%%%%%%%

To build up our intuition about the type of dual answer we can get it is instructive to consider the particular case of massive fermions, where the potential $V(\bar{\psi}\psi)$ just leads to a mass term. Let us consider integrating over the fermionic degrees of freedom before we integrate out the gauge field. In the large mass limit we have:

\bea
Z^F[\Lambda_\mu^a]_{m\to \infty}&=&\int DA_\mu^a D\bar{\psi}D\psi\exp\bigg[\int d^3x \left(\bar{\psi}(i\slashed{\partial}+m +i\slashed{A})\psi + \epsilon^{\mu\nu\rho}F_{\mu\nu}^a \Lambda_\rho^a\right)-S^{CS}_k\bigg]\nonumber\\
&\approx &\int DA_\mu^a\exp\left(S^{CS}_{\Delta k}(A)+ \int d^3x \epsilon^{\mu\nu\rho}F_{\mu\nu}^a \Lambda_\rho\right),
\eea
where the Chern-Simons level is  $\Delta k=k_1-k$ with $k_1=N\tr \mathbbm{1}_2=2N$, where $\mathbbm{1}_2$ is the identity matrix in the space of Dirac gamma matrices. We have used that in the large mass limit the integration over fermions leads to a Chern-Simons action for the corresponding gauge field. This is a fairly standard computation which can be found, for example, in \cite{Dunne:1998qy}. We provide a brief review of the calculation in appendix A  for self-completeness and to fix out notation.

The next step is to integrate over the gauge field. With this aim, we find it convenient to take the light cone gauge
\be
A_- =\frac{1}{\sqrt{2}}\left(A_1+A_2\right)=0.
\ee
In the line-cone gauge the gluon self-energy vanishes and the Chern-Simons actions becomes
\be
S^{CS}_k=\frac{k}{4\pi}\int d^3 x A_+^a\partial_-A_3^a.
\ee

The path integral takes the following form:
\bea
Z^F[\Lambda_\mu^a]_{m\to \infty}&=& \int DA_+^a DA_3^a \exp\bigg[\int d^3x \left(A_+^a \left(\frac{\Delta k}{2\pi}\delta^{ab}\partial_- - f^{ab}{}_c \Lambda^c_-\right)A_3^b+ A_+^a\cf_{-3}^a + A_3^a\cf_{+-}^a\right)\bigg], \nonumber
\eea

where
\be
\cf_{\mu\nu}^a=\partial_\mu \Lambda^a_\nu -\partial_\nu \Lambda_\mu^a.
\ee
The integration of $A_+^a$ leads to the following equation for $A_3^a$:
\be
\left(\frac{\Delta k}{2\pi}\delta^{ab}\partial_- - f^{ab}{}_c \Lambda_-^c\right)A_3^b+{\cal F}_{-3}^a=0.
\ee
Denoting $N^{ab}$ the inverse of the operator multiplying $A_3^b$ in the equation above, i.e.,
\be
\label{Eq:Nab1}
\left(\frac{\Delta k}{2\pi}\delta^{ac}\partial_- - f^{ac}{}_d \Lambda_-^d\right)N_{cb}=\delta^a_b,
\ee
the final result for the partition function of the dual theory in this limit is:

\bea
Z^F[\Lambda_\mu^a]_{m\to \infty}&=&\int DA_\mu^a\exp\left(S^{CS}_{\Delta k}(A)+ \epsilon^{\mu\nu\rho}F_{\mu\nu}^a \Lambda_\rho\right), \\
&=& \int DA_+^a DA_3^a \exp\left(A_+^a \left(\frac{\Delta k}{2\pi} \delta^{ab}\partial_- - f^{ab}{}_c \Lambda^c_-\right)A_3^b+ A_+^a\cf_{-3}^a + A_3^a\cf_{+-}^a\right) \nonumber
\eea

The partition function turns out to be
\be
Z^F[\Lambda_\mu]_{m\to \infty}= \int D\Lambda_\mu \left(\det(\frac{\Delta k}{2\pi} \delta^{ab}\partial_- - f^{ab}{}_c \Lambda_-^c)\right)^{-1} \exp\left(-\int d^3x N^{ab}\cf^a_{-3}\cf^b_{+-}\right).
\ee

A few comments about this answer are in order.  The above expression has been obtained in the large mass limit and therefore it is expected to capture phenomena in the long wave length range.  One might object that the action does not look appropriately non Abelian, in the sense that $\cf_{\mu\nu}^a=\partial_\mu \Lambda_\nu^a -\partial_\nu\Lambda_\mu^a$ does not contain the expected commutator term.  The answer to this seeming puzzle  is simply that the natural object to dualize the action would be a Kalb-Ramond 2-form field ($F_{\mu\nu}^a \Lambda^a{}^{\mu\nu}$)  we are simply rewriting the action in terms of the dual field $\L^a_\rho =\epsilon_{\rho}{}^{\mu\nu}\L^a_{\mu\nu}$.

%%%%%%%%%%%%%%%%%%%%%%%%%%%%%%%%%%%%%
\subsection{Light-cone gauge and the large-$N$ limit}
%%%%%%%%%%%%%%%%%%%%%%%%%%%%%%%%%%%%%%%%%%
We now consider integrating the fermionic degrees of freedom in the large $N$ limit. First, we implement the integration over the gauge fields using the light-cone gauge: $A_-=0$. The action reduces to
\bea
S&=&\int d^3x \bigg[\bar{\psi}\gamma^\mu \partial_\mu \psi +A_+^aJ_-^ a+A_3^aJ_3^a+ \frac{k}{2\pi}A^a_+\partial_- A_3^a +V(\bar{\psi}\psi) \nonumber \\
&+&A_+^a\partial_-\L_3^a -A_3^a \partial_-\L_+^a-A_+^a\partial_3 \L_-^a+A_3^a\partial_- \L_-^a + f_{abc}A_3^bA_+^c \L_-^a\bigg], \nonumber \\
&=&\int d^3x \bigg[\bar{\psi}\gamma^\mu \partial_\mu \psi +V(\bar{\psi}\psi)  \nonumber \\
&+&A_+^a\left(J_-^ a+ \frac{k}{2\pi}A^a_+\cf^a_{-3}+ f_{abc}\Lambda^b_- A_3^c\right)  \nonumber \\
&+&A_3^a(J_3^a +\cf_{+-}^a)\bigg],
\eea

where
\be
J^a_\mu=-i\bar{\psi}\gamma_\mu T^a \psi.
\ee

The action is linear in $A_+$ (it is also linear in $A_3$) and we will integrate it out by using its equation of motion:
\be
\label{Eq:A-Fermions}
\frac{k}{2\pi}\partial_- A_3^a +J_-^a +\partial_-\L_3^a -\partial_3 \L_-^a+f_{abc}\L_-^bA_3 ^c=0.
\ee
One limit of this equation $(\L\equiv 0)$ was treated in \cite{Jain:2012qi} we will return to this limit repeatedly as a source of technical and conceptual intuition. We rewrite the equation as:
\be
\label{Eq:Nab2}
\left(\frac{k}{2\pi}\delta^{ab}\partial_- - f^{ab}{}_c \Lambda_-^c\right)A^b_3=J_-^a +\cf_{-3}^a.
\ee
Let us formally define the solution to be
\be
A_3^a=N^{ab}(J_-^b+\cf_{-3}^b),
\ee
where $N^{ab}$ is the same tensor that appeared already in Eq. (\ref{Eq:Nab1}) with the small difference of the Chern-Simons level.  Substituting in the action we arrive at
\bea
\label{Eq:FermionsIntermediate}
S&=& \int d^3x \bigg[ \bar{\psi}\left(\gamma^\mu \partial_\mu -i N^{ab}\cf_{-3}^b\gamma_3 T^a +i N^{ab}\cf_{-+}^a \gamma_- T^b\right)\psi +V(\bar{\psi}\psi) \nonumber \\
&-& N^{ab}\bar{\psi}\gamma_-T^b \psi \bar{\psi}\gamma_3 T^a \psi- N^{ab} \cf_{-+}^a\cf_{-3}^b\bigg]
\eea
The dual theory takes the following form:
\bea
\label{Eq:FermionsFormal}
Z^F[\Lambda_\mu^a]&=& \int D\bar{\psi}D\psi \det \left(\frac{k}{4\pi}\delta^{ab}\partial_- - f^{ab}{}_c \Lambda_-^c\right)^{-1}\exp\left(\bar{\psi}\slashed{\partial}\psi +V(\bar{\psi}\psi) + A_3^a (J_3^a+\cf_{+-}^a)\right) \nonumber \\
&=& \int D\bar{\psi}D\psi \exp\bigg(\bar{\psi}\left(\gamma^\mu \partial_\mu -i N^{ab}\cf_{-3}^b\gamma_3 T^a +i N^{ab}\cf_{-+}^a \gamma_- T^b\right)\psi +V(\bar{\psi}\psi) \nonumber\\
& -&   N^{ab}\bar{\psi}\gamma_-T^b \psi \bar{\psi}\gamma_3 T^a \psi\bigg) %\nonumber \\
\times  \det \left(\frac{k}{4\pi}\delta^{ab}\partial_- - f^{ab}{}_c \Lambda_-^c\right)^{-1}\exp\left( - N^{ab} \cf_{-+}^a\cf_{-3}^b\right).\nonumber\\
& &
\eea
Before proceeding to integrate over fermions we already notice the appearance, in the last  line in (\ref{Eq:FermionsFormal}), of part of the structure that showed up in the large mass limit. In particular, the last line is proving to be a recurrent structure of the effective action for the field $\Lambda_\mu^a$ in various limits.
%Another interesting representation of  $Z^F[\Lambda_\mu^a]$ can be given by using the
%following integral
%\bea
%e^{-N^{ab}J_-^{a}J_3^{b}}=\frac{1}{\sqrt{Det(N^{ab})}}\int d \eta_3^c  d \eta_{-}^d e^{\int d^3x (-\eta_3^a\eta_-^b (N^{ab}) ^{-1}+\eta_3^a J_-^a+\eta_-^a J_3^a)}
%\eea
%So using the above integral representation we get
%\bea
%Z^F[\Lambda_\mu^a]=& \int D\bar{\psi}D\psi e^{ - \int d^3x \bigg[ \bar{\psi}\left(\gamma^\mu \partial_\mu -i N^{ab}\cf_{-3}^b\gamma_3 T^a +i N^{ab}\cf_{-+}^a \gamma_- T^b\right)\psi \bigg]}\times
%\nonumber \\
%& e^{- \int d^3x \bigg[ N^{ab}\bar{\psi}\gamma_-T^b \psi \bar{\psi}\gamma_3 T^a \psi- N^{ab} \cf_{-+}^a\cf_{-3}^b\bigg]}\nonumber \\
%=&\frac{1}{\sqrt{Det(N^{ab})}} e^{-\int d^3x N^{ab} \cf_{-+}^a\cf_{-3}^b}\int d \eta_3^c  d \eta_{-}^d Det(\gamma^{\mu}\partial_\mu-i (\eta_3^a \gamma_-+\eta_-^a \gamma_3)T^a )
%\nonumber\\
%&e^{-\int d^3x N^{ab^{-1}\lbrack(\eta_3^a-N^{ac} \cf_{+-}^c)(\eta_-^b-N^{be} \cf_{-3}^e)+(\eta_3^b-N^{bc} \cf_{+-}^c)(\eta_-^a-N^{ae} \cf_{-3}^e)\rbrack}}
%\eea
%where in reaching the last expression we have made some shifts with unit Jacobian in the auxiliary fields $\eta_3^a,\eta_-^a$.

We have a formal expression  for the dual action in Eq. (\ref{Eq:FermionsFormal}), to achieve a better understanding  and to prepare the groundwork for the large $N$ limit, it is convenient to study in more detail the group structure for the operator $N^{ab}$ as follows. The natural decomposition of any algebra valued two-tensor  should  be:
\bea
N^{ab}=\delta^{ab}\mathcal{N}_0+ i f^{abc}\mathcal{N}_1^c+d^{abc}\mathcal{N}_2^c.
\eea
Using the fact that
\bea
N^{ab}\left(\frac{k}{2\pi}\delta^{bd} \partial_- -f^{bdc}\L_-^c\right)A_3^d=A_3^a
\eea
we find
\bea
\label{Eq:N0}
\mathcal{N}_0=\frac{2 \pi}{k}\,\, \frac{1}{ \partial_-}-\frac{1}{N}\frac{2 \pi}{k}\,\, \frac{1}{ \partial_-} i \mathcal{N}_1^a\L_-^a
\eea
%and $\mathcal{N}_1^a,\mathcal{N}_2^a$  satisfy the following coupled system of equations
%\bea
%\frac{i k N \mathcal{N}_1^e \partial_-}{2 \pi}-\mathcal{N}_0 N \L_-^e-i f^{abc}f^{bdl}f^{ead}\L_-^c\mathcal{N}_1^l-f^{abc}d^{bdl}f^{ead}\L_-^c\mathcal{N}_2^l=0,
%\eea
%\bea
%\frac{ik (N^2-4) \mathcal{N}_2^e \partial_-}{2 \pi N}-i f^{abc}f^{bdl}d^{ead}\L_-^c\mathcal{N}_1^l-f^{abc}d^{bdl}d^{ead}\L_-^c\mathcal{N}_2^l=0.
%\eea
Using the identity
\bea
f^{dan}f^{abc}f^{bde}=N f^{nce},
\eea
we find coupled system of equations for $\mathcal{N}_1^a,\mathcal{N}_2^a$,
\bea
\label{Eq:N1}
\frac{i k  \mathcal{N}_1^e \partial_-}{2 \pi}-i  f^{elc}\L_-^c\mathcal{N}_1^l=\mathcal{N}_0  \L_-^e+ d^{elc}\L_-^c\mathcal{N}_2^l,
\eea
\bea
\label{Eq:N2}
\frac{ k (N^2-4) \mathcal{N}_2^e \partial_-}{2 \pi N^2}-  f^{elc}\L_-^c\mathcal{N}_2^l= i   d^{elc}\L_-^c\mathcal{N}_1^l.
\eea
%\bea
%\frac{i k  \mathcal{N}_1^e \partial_-}{2 \pi}-\mathcal{N}_0  \L_-^e-i  f^{elc}\L_-^c\mathcal{N}_1^l-  d^{elc}\L_-^c\mathcal{N}_2^l=0,
%\eea
%\bea
%\frac{i k (N^2-4) \mathcal{N}_2^e \partial_-}{2 \pi N^2}+ i   d^{elc}\L_-^c\mathcal{N}_1^l-  f^{elc}\L_-^c\mathcal{N}_2^l=0.
%\eea

These equations should be taken in the formal sense due to the presence of $\partial_-$ . We will discuss other aspects of $N^{ab}$, or equivalentely of the integro-differential equation for $A_3^a(\Lambda)$ in appendix \ref{Sec:MomentumSpace}.  Now, in this basis the quartic fermion term in the  previous action Eq. (\ref{Eq:FermionsIntermediate}) becomes :
\bea
\label{Eq:FermionsIntermediate2}
S&=& \int d^3x \bigg[ \bar{\psi}\bigg(\gamma^\mu \partial_\mu \nonumber\\ 
&-&i (\delta^{ab}\mathcal{N}_0+ i f^{abc}\mathcal{N}_1^c+d^{abc}\mathcal{N}_2^c)(\cf_{-3}^b\gamma_3 T^a + \cf_{-+}^a \gamma_- T^b)\bigg)\psi +V(\bar{\psi}\psi) \nonumber \\
&-& (\delta^{ab}\mathcal{N}_0+ i f^{abc}\mathcal{N}_1^c+d^{abc}\mathcal{N}_2^c)(\bar{\psi}\gamma_-T^b \psi \bar{\psi}\gamma_3 T^a \psi-  \cf_{-+}^a\cf_{-3}^b)\bigg]
\eea

%\bea
%S&=& \int d^3 \bigg[ \bar{\psi}\left(\gamma^\mu \partial_\mu -i N^{ab} \cf_{-3}^b\gamma_3 T^a
 %+i N^{ab}\cf_{-+}^a \gamma_- T^b\right)\psi +V(\bar{\psi}\psi) \nonumber \\
%&-&(-\mathcal{N}_0\bar{\psi}\gamma_-\psi\bar{\psi}\psi-\frac{1}{N}\mathcal{N}_0\bar{\psi}\gamma_-\psi\bar{\psi}\gamma_3\psi+\bar{\psi}\gamma_-\mathcal{N}_1\psi\bar{\psi}\psi-\bar{\psi}\mathcal{N}_1\psi\bar{\psi}\gamma_-\psi  \nonumber\\
%&+&\frac{1}{2}\bar{\psi}\gamma_-\mathcal{N}_2\psi\bar{\psi}\psi
%-\frac{1}{2}\bar{\psi}\mathcal{N}_2\psi\bar{\psi}\gamma_-\psi-\frac{1}{N}\bar{\psi}\gamma_3\mathcal{N}_2\psi\bar{\psi}\gamma_-\psi-\frac{1}{N}\bar{\psi}\gamma_-\mathcal{N}_2\psi\bar{\psi}\gamma_3\psi
%\nonumber\\
%&-& N^{ab} \cf_{-+}^a\cf_{-3}^b)\bigg],\nonumber\\
%\eea
where $\mathcal{N}_2=\mathcal{N}_2^a T^a,\mathcal{N}_1=\mathcal{N}_1^a T^a$. 

First, observe that the terms which have explicit coefficient of  $\frac{1}{N}$  vanish in the $N\to\infty$ limit. Next,  we exploit the fact  that $\mathcal{N}_0\sim {\cal O}(\frac{1}{k})$. We also observe that from the coupled equations for $\mathcal{N}_1$ and $\mathcal{N}_2$ it follows that   $\mathcal{N}_1$ is suppressed by powers of  $\frac{1}{k^2}$ or higher powers of $\frac{1}{k}$, and   that $\mathcal{N}_2$ is suppressed by $\frac{1}{k^3}$ or higher powers of $\frac{1}{k}$.
Therefore, all the terms in the actions involving $\L^a_{\mu}$, except the universal term
$ N^{ab} \cf_{-+}^a\cf_{-3}^b$,  are subleading in the limit $N\to\infty, k\to\infty$, with  $\frac{N}{k}$  fixed. This universal term remains untouched because we are not integrating over $\L^a_{\mu}$.

%Therefore clearly the four fermion terms $\bar{\psi}\gamma_-\mathcal{N}_1\psi\bar{\psi}\psi$,$\bar{\psi}\mathcal{N}_1\psi\bar{\psi}\gamma_-\psi$,
%$\bar{\psi}\gamma_-\mathcal{N}_2\psi\bar{\psi}\psi$ \\ and $\bar{\psi}\mathcal{N}_2\psi\bar{\psi}\gamma_-\psi$  will not give contributions to the singlet sector of the theory in the Large-N limit.
%This will be explained more explicity in the appendix.
%\bea
%\bar{\psi}( \mathcal{N}_0\cf_{-3}\gamma_3+\lbrack \cf_{-3},\mathcal{N}_1\rbrack \gamma_3+\lbrack \mathcal{N}_1,\cf_{-3}\rbrack \gamma_3+\{ \cf_{-3},\mathcal{N}_1\} \gamma_3+\{\mathcal{N}_1,\cf_{-3}\} \gamma_3+
%\nonumber\\
%\mathcal{N}_0\cf_{+-}\gamma_-+\lbrack \cf_{+-},\mathcal{N}_1\rbrack \gamma_-+\lbrack \mathcal{N}_1,\cf_{+-}\rbrack \gamma_-+\{ \cf_{+-},\mathcal{N}_1\} \gamma_-+\{\mathcal{N}_1,\cf_{+-}\} \gamma_-)\psi
%\eea

The main obstruction to a full integration of the fermionic fields remains the term quartic in fermions.
The starting identity in the large-$N$ integration is a simple extension of the Hubbard-Stratonovich identity \cite{Jain:2012qi}.
Here we assume the bi-local structure $\bar{\psi}^i(x)\psi^i(y)$ as discussed in detail in \cite{Jain:2012qi} which highlights the connection to higher spin fields \cite{Koch:2010cy}; we also do not write explicitly integration over spacetime variables. However, we assume both implicitly. 

\bea\label{Eq:hubbard}
1&=& \int D\alpha_{0}D\alpha_{-}\delta \left(\alpha_{0}-\frac{1}{2N}\bar{\psi}\psi\right)
\delta\left(\alpha_{-}-\frac{1}{2N}\bar{\psi}\gamma_-\psi\right) \nonumber\\
&=& \int D\alpha_{0}D\alpha_{-}D\mu_0 D\mu_+  \exp\left(2i\mu_0\left(\alpha_{0}-\frac{1}{2N}\bar{\psi}\psi\right)
+2i\mu_+\left(\alpha_{-}-\frac{1}{2N}\bar{\psi}\gamma_-\psi)\right)\right).\nonumber\\
\eea
The partition function is given by
\bea
Z&=&\int \mathcal{D}\bar{\psi}\mathcal{D}\psi e^{-\tr\log(\frac{k}{4\pi}\delta^{ab}\partial_- - f^{ab}{}_c \Lambda_-^c)}\exp\bigg[  \int \frac{d^3q}{(2\pi)^3}\bar{\psi}(-q)\gamma^{\mu}q_{\mu}\psi(q)
\nonumber\\
&+&N \int \frac{d^3P}{(2\pi)^3}\frac{d^3p_1}{(2\pi)^3}\frac{d^3p_2}{(2\pi)^3}\frac{8\pi i N}{k (p_1-p_2)_-}\xi_-(P,p_1)\xi_I(-P,p_2) \bigg]
\eea
Inserting the identity back in the partition function it becomes
\bea
Z&=&\int \mathcal{D}\alpha_0 \mathcal{D}\alpha_- \mathcal{D}\mu_0 \mathcal{D}\mu_+ \exp\bigg(i\int (2\mu_+.\alpha_-+2\mu_0.\alpha_0)\bigg)\times\nonumber\\
& &\exp\bigg(-N\int \frac{d^3P}{(2\pi)^3}\frac{d^3p_1}{(2\pi)^3}\frac{d^3p_2}{(2\pi)^3}\frac{8\pi i N}{k (p_1-p_2)_-}\alpha_-(P,p_1)\alpha_0(-P,p_2)\bigg)\times\nonumber\\
& &\int \mathcal{D}\bar{\psi}\mathcal{D}\psi \exp\left({-\tr\log(\frac{k}{4\pi}\delta^{ab}\partial_- - f^{ab}{}_c \Lambda_-^c)}\right)\times \\
& &\exp\bigg(\int \frac{d^3qd^3P}{(2\pi)^6}
\bar{\psi}(-\frac{P}{2}-q)(i\gamma^{\mu}q_{\mu}(2\pi)^3\delta^3(P)-i\frac{\gamma_- \mu_+}{N}-\frac{i \mu_0}{N}-i\frac{\gamma_-\cf_{+-}}{k P_-}-i\frac{\gamma_3\cf_{-3}}{k P_-}) \psi(-\frac{P}{2}+q)\bigg)\nonumber
\eea
rescaling
\bea
\mu_0=N \tilde{\mu}_0,\quad \tilde{\mu}_+=N \mu_+
\eea
In the large N limit the path integral is dominated by the saddle point solutions and assuming Poincare invariance on these saddle point solutions we get
\bea
\langle\alpha_0(P,p_1)\rangle=(2\pi)^3\delta^3(P)\alpha_0(p_1),\quad  \langle\alpha_-(P,p_1)\rangle=(2\pi)^3\delta^3(P)\alpha_-(p_1)\nonumber\\
\langle i \mu_0(P,p_1)\rangle=(2\pi)^3\delta^3(P)\Sigma_0(p_1),\quad \langle i \mu_-(P,p_1)\rangle=(2\pi)^3\delta^3(P)\Sigma_-(p_1)
\eea
At this vacuum configuration of the singlet fields the partition function becomes
\bea
Z&=&\int \mathcal{D}\alpha_0 \mathcal{D}\alpha_- \mathcal{D}\mu_0 \mathcal{D}\mu_+ exp\bigg(i\int (2\mu_+.\alpha_-+2\mu_0.\alpha_0)\bigg)\times\nonumber\\
& &\exp\bigg(-V N\int \frac{d^3p_1}{(2\pi)^3}\frac{d^3p_2}{(2\pi)^3}\frac{8\pi i N}{k (p_1-p_2)_-}\alpha_-(p_1)\alpha_0(p_2)\bigg)\times%\nonumber\\
%& & 
e^{-\tr\log(\frac{k}{4\pi}\delta^{ab}\partial_- - f^{ab}{}_c \Lambda_-^c)}\times
\nonumber\\
& &\exp\bigg(-\int \frac{d^3qd^3P}{(2\pi)^6}
\bar{\psi}(-\frac{P}{2}-q)(i\gamma^{\mu}q_{\mu}+\gamma_- \Sigma_+ +\Sigma_0)(2\pi)^3\delta^3(P) \psi(-\frac{P}{2}+q)\bigg)\nonumber\\
&= &\int \mathcal{D}\alpha_0 \mathcal{D}\alpha_- \mathcal{D}\mu_0 \mathcal{D}\mu_+ exp\bigg(i\int (2\mu_+.\alpha_-+2\mu_0.\alpha_0)\bigg)\times\nonumber\\
& &\exp\bigg(-V N\int \frac{d^3p_1}{(2\pi)^3}\frac{d^3p_2}{(2\pi)^3}\frac{8\pi i N}{k (p_1-p_2)_-}\alpha_-(p_1)\alpha_0(p_2)\bigg)\times%\nonumber\\
%& &
 e^{-\tr\log(\frac{k}{4\pi}\delta^{ab}\partial_- - f^{ab}{}_c \Lambda_-^c)}\times
\nonumber\\
& &\exp\bigg(-V N \int \frac{d^3q}{(2\pi)^3}\log\det(i \gamma^{\mu}q_{\mu}+\gamma_- \Sigma_+ +\Sigma_0\bigg)
\eea
and
\bea
-i \int \tilde{\mu}.\alpha=V\int\frac{d^3q}{(2\pi)^3}\bigg( - 2 \Sigma_+(q)\alpha_-(q)- 2 \Sigma_0(q)\alpha_0(q)\bigg)
\eea
Where $V\equiv (2\pi)^3\delta^3(0)$ and we have used the following identity for fermionic Gaussian integration
\bea
\int \mathcal{D}\psi\mathcal{D}\bar{\psi}e^{-\int \frac{d^3p}{(2\pi)^3}\bar{\psi}(-p)A(p)\psi(p)}=e^{V\int \frac{d^3p}{(2\pi)^3}\log\det A(p)}
\eea. 

All these pieces combine to give the effective action
\bea\label{Eq:saddlepsi}
S_{eff}&=&N V\bigg(- \int \frac{d^3q}{(2\pi)^3}\log\det(i \gamma^{\mu}q_{\mu}+\gamma_{-} \Sigma_{+}
+\Sigma_{0})
\nonumber\\
&+&\int \frac{d^3p_1}{(2\pi)^3}\frac{d^3p_2}{(2\pi)^3}\frac{8\pi i N}{k (p_1-p_2)_-}\alpha_-(p_1)\alpha_0(p_2) )
\nonumber\\
&+&\int\frac{d^3q}{(2\pi)^3}( - 2 \Sigma_+(q)\alpha_-(q)- 2 \Sigma_0(q)\alpha_0(q))
\bigg)
\eea
 
Varying $S_{eff}$ action with respect to $\alpha_-,\alpha_0$ and putting the resulting equations back
we get the following gap equations
\bea
&\Sigma_{F,+}(p)=-2 \pi i \lambda \int \frac{d^3 q}{(2\pi)^3}\frac{1}{(p-q)_-}\tr(\frac{1}{i\gamma^{\mu}q_{\mu}+\Sigma_{F}(q)})\nonumber\\
&\Sigma_{F,I}(p)=2 \pi i \lambda \int \frac{d^3 q}{(2\pi)^3}\frac{1}{(p-q)_-}\tr(\frac{\gamma_-}{i\gamma^{\mu}q_{\mu}+\Sigma_{F}(q)})
\eea
and the full dual effective action
\bea
\label{Eq:FermionsFinalDual}
S_{eff}[\L] & = & N V\bigg[\int \frac{d^3 q}{(2\pi)^3}\left\{ -\log \det (i\gamma^{\mu}q_{\mu}+\Sigma_{F}(q))+\frac{1}{2}\frac{\tr\Sigma_{F}(q)}{i \gamma^{\mu}q_{\mu}+\Sigma_{F}(q)}\right\}\bigg]\nonumber\\
&+& S[\L]
\eea
where
\bea
S[\L]=-\int d^3x N^{ab} \cf_{-+}^a\cf_{-3}^b -\tr\log\left(\frac{k}{4\pi}\delta^{ab}\partial_- - f^{ab}{}_c \Lambda_-^c\right)\label{eq:sl}
\eea
 which does not contribute to the gap equations in the Large N limit.

The precise evaluation of the first line in Eq. \ref{Eq:FermionsFinalDual}  which is the planar free energy  has been the main result of a series of works starting with \cite{Giombi:2011kc} and elaborated upon in, for example, \cite{Aharony:2012ns,GurAri:2012is,Jain:2012qi,Jain:2013gza}. The main conclusion is that $\Sigma_F(q)$ can be computed exactly in the large $N$ limit and it corresponds to the self-energy of the fermions. The large-$N$ approach that we have followed here has also been corroborated by a Schwinger-Dyson equation perspective in some of the works cited above.

It is worth noting that the large mass limit of the previous subsection can be recovered here as well before integrating out the gauge field. The gap equation for $\alpha_0$ shows that $\mu_0$ becomes the mass as read off from the potential. A diagrammatic argument leads to the action for $\mu_+, \mu_3$ to be a Chern-Simons action. The other gap equations identify $A_+, A_3$ with $\mu_+ $ and $\mu_3$. We thus recover precisely the same result as in the previous subsection.

%%%%%%%%%%%%%%%%%%%%%%%%%%%%%%%%%%%%%%%%%%%%%%%%%%%%%%%%%%%%%%%%%%%%%%%%%%%%%%%%%%%%%%%%
\section{Duality for Bosonic Vector Models }\label{Sec:Boson}
We begin with the following action corresponding to a single fundamental boson
\be
S_\phi=\int d^3x \left(\frac{1}{2}\partial_\mu \bar{\phi}\partial^\mu\phi+ U(\bar{\phi}\phi)\right).
\ee
We proceed to gauge the global $U(N)$ symmetry by introducing a gauge field $A_\mu$ and adding a Lagrange multiplier $\L$ to the action that enforces the vanishing of the field strength

\be
S=\int d^3x \bigg(D_\mu \bar{\phi} D^\mu \phi + U(\bar{\phi}\phi)+\epsilon^{\mu\nu\rho}F_{\mu\nu}\L_\rho\bigg),
\ee
where

\be
D_\mu \phi = \partial_\mu \phi^m -iA_\mu^a T_a  \phi, \qquad D_\mu \bar{\phi} = \partial_\mu \bar{\phi}^m +iA_\mu^a T_a  \bar{\phi}.
\ee
 With the same motivation of the previous section we add a Chern-Simons term, the full starting action becomes
\bea
S&=& \int d^3x \bigg[\partial_\mu\bar{\phi}\partial^\mu\phi +A_\mu^aJ_\mu^a+ A_\mu^a A_\mu^b \bar{\phi}T^a T^b \phi %\nonumber \\
+ \epsilon^{\mu\nu\rho}F_{\mu\nu}\L_\rho+ U(\bar{\phi}\phi)\nonumber \\
&+&\frac{ik}{4\pi}\epsilon^{\mu\nu\rho}\tr \left(A_\mu  \partial_\nu A_\rho  -\frac{2i}{3}A_\mu  A_\nu A_\rho \right) \bigg].
\eea
where
\be
\label{Eq:BosonCurrent}
J_{\mu}^a=i(\bar{\phi}T^a \partial_\mu \phi -(\partial_\mu \bar{\phi})T^a \phi).
\ee
As in the previous sections our aim is to integrate out the bosonic field $\phi$ and the gauge field $A_\mu^a$ leaving a dual action for $\Lambda_\mu^a$. We first proceed to integrate out the gauge field. The most efficient way is using the light-cone gauge ($A_-=0$). In the light-cone gauge the action becomes
\bea
S&=& \int d^3x \bigg[-\bar{\phi}\left(2\partial_+\partial_-+\partial_3^2\right)\phi +A_+^aJ_-^a+A_3^a J_3^a + A_3^a A_3^b \bar{\phi}T^a T^b \phi \nonumber \\
&-&\partial_-A^a_+\L^a_3+\partial_- A^a_3 \L^a_+  + (\partial_3A_+^a -\partial_+A_3^a +f_{abc}A_3^b A_+^c)\L_-^a+\frac{k}{2\pi}A_+^a\partial_- A_3^a\bigg]
\eea
where
\be
J_{\mu}^a=i(\bar{\phi}T^a \partial_\mu \phi -(\partial_\mu \bar{\phi})T^a \phi).
\ee

The action is linear in $A_+^a$ whose corresponding equation of motion leads to
\be
\label{Eq:A-Bosons}
\left(\frac{k}{2\pi}\delta^{ab}\partial_- -f^{ab}{}_c\L_-^c\right)A_3^b+J_-^a +\cf_{-3}^a=0.
\ee
We assume that the solution of this equation is of the form
\be
A_3^a =N^{ab}(J_-^b +\cf_{-3}^b),
\ee
where $N^{ab}$ is the, by now ubiquotous, formal  tensor which appeared in Eqns. (\ref{Eq:Nab1}) and (\ref{Eq:Nab2}). Note that in this case the Chern-Simons level enters exactly as in the case of  as in Eq. (\ref{Eq:Nab2}).  Substituting this solution into the action one arrives at

\bea
S&=&-\bar{\phi}\left(2\partial_+\partial_-+\partial_3^2\right)\phi  \nonumber \\
&+&N^{ab}(J_-^b+\cf_{-3}^b)(J_3^a+\cf_{+-}^a) \nonumber \\
&+& N^{ac}N^{bd}(J_-^c+\cf_{-3}^c)(J_-^d+\cf_{-3}^d)\bar{\phi}T^aT^b\phi 
\eea
Similar to the fermionic case, to gain intuition into the dual action we expand the tensorial structures in $N^{ab}$ by substituting
\bea
N^{ab}=\delta^{ab}\mathcal{N}_0+ i f^{abc}\mathcal{N}_1^c+d^{abc}\mathcal{N}_2^c.
\eea
to get
\bea
S&=&-\bar{\phi}\left(2\partial_+\partial_-+\partial_3^2\right)\phi  \nonumber \\
&+&(\delta^{ab}\mathcal{N}_0+ i f^{abc}\mathcal{N}_1^c+d^{abc}\mathcal{N}_2^c)(J_-^b+\cf_{-3}^b)(J_3^a+\cf_{+-}^a) \nonumber \\
&+& (\delta^{ac}\mathcal{N}_0+ i f^{acc_1}\mathcal{N}_1^{c_1}+d^{acc_1}\mathcal{N}_2^{c_1})(\delta^{bd}\mathcal{N}_0+ i f^{bdd_1}\mathcal{N}_1^{d_1}+d^{bdd_1}\mathcal{N}_2^{d_1})\nonumber\\
&\times&(J_-^c+\cf_{-3}^c)(J_-^d+\cf_{-3}^d)\bar{\phi}T^aT^b\phi 
\eea
%To know the $N$ dependence of every term it's feasible to be more explicit
%\bea
%&(\delta^{ab}\mathcal{N}_0+ i f^{abc}\mathcal{N}_1^c+d^{abc}\mathcal{N}_2^c)(J_-^b+\cf_{-3}^b)(J_3^a+\cf_{+-}^a)\nonumber\\ &=\mathcal{N}_0(J_-^a+\cf_{-3}^a)(J_3^a+\cf_{+-}^a) 
%+ i f^{abc}\mathcal{N}_1^c(J_-^a+\cf_{-3}^a)(J_3^a+\cf_{+-}^a)\nonumber\\
%&+d^{abc}\mathcal{N}_2^c(J_-^a+\cf_{-3}^a)(J_3^a+\cf_{+-}^a)
%\eea
%and
%\bea
%&(\delta^{ac}\mathcal{N}_0+ i f^{acc_1}\mathcal{N}_1^{c_1}+d^{acc_1}\mathcal{N}_2^{c_1})(\delta^{bd}\mathcal{N}_0+ i f^{bdd_1}\mathcal{N}_1^{d_1}+d^{bdd_1}\mathcal{N}_2^{d_1})\nonumber\\
%&= \delta^{ac}\delta^{bd}\mathcal{N}_0^2+i\delta^{ac}f^{bdd_1}\mathcal{N}_1^{d_1}\mathcal{N}_0+\delta^{ac}d^{bdd_1}\mathcal{N}_2^{d_1}\mathcal{N}_0+i\delta^{bd}f^{acc_1}\mathcal{N}_1^{c_1}\mathcal{N}_0\nonumber\\
%&-f^{acc_1}f^{bdd_1}\mathcal{N}_1^{c_1}\mathcal{N}_1^{d_1}+i f^{acc_1}d^{bdd_1}\mathcal{N}_1^{c_1}\mathcal{N}_2^{d_1}+\delta^{bd}d^{acc_1}\mathcal{N}_2^{c_1}\mathcal{N}_0\nonumber\\
%&+i d^{acc_1}f^{bdd_1}\mathcal{N}_2^{c_1}\mathcal{N}_1^{d_1}+d^{acc_1}d^{bdd_1}\mathcal{N}_2^{c_1}\mathcal{N}_2^{d_1}
%\eea
%in the bosonic action, it is easy to see that the only contribution to the singlet sector comes from
%$\delta^{ab}\mathcal{N}_0$ part of $N^{ab}$. Just like the fermonic counterpart all the terms involving $\mathcal{N}_1^c$ and $\mathcal{N}_2^c$ will not contribute to the dual action for $\L^a$ in the large $N$ limit.

It is possible to integrate the bosons in the large $N$ limit by using an extended version of Hubbard-Stratonovich. We implement the Hubbard-Stratonovich identity at the level of path integral as
\be\label{Eq:hubbard2}
1=\int D\mu D\alpha \exp\bigg(i \mu (\alpha -\frac{1}{N}\bar{\phi} \phi)\bigg).
\ee
Here we assume the bi-local structure $\bar{\phi}^i(x)\phi^i(y)$ as discussed in detail in \cite{Jain:2012qi} which highlights the connection to higher spin fields \cite{Koch:2010cy}; we simply do not write explicitly the dependence in the coordinates.

After repeating the steps we did for fermonic action we get the following result for the gap equation
for the bosonic theory.
\bea\label{Eq:Sigmaphi}
\Sigma_B(p)&=&\int\frac{d^3 q_1}{(2\pi)^3}\frac{d^3 q_2}{(2\pi)^3}\bigg[ C_2(p,q_1,q_2)+C_2(q_1,p,q_2)+C_2(q_1,q_2,p) \bigg]\frac{1}{q_1^2+\Sigma_B(q_1)}\frac{1}{q_2^2+\Sigma_B(q_2)}\nonumber\\
& & 
\eea
where
\bea
C_2(p_1,p_2,p_3)=4 \pi^2 \lambda^2 \left(\frac{(p_1+p_3)_-(p_2+p_3)_-}{(p_1-p_3)_-(p_2-p_3)_-}+(coupling[(\bar{\phi}\phi)^3])\right)
\eea
 and the effective action
\bea
\label{Eq:BosonsFinalDual}
S_{eff}[\L]&=& N V\bigg[\int \frac{d^3 q}{(2\pi)^3}\left\{ \log (q_s^2+q_3^2+\Sigma_{B}(q))-\frac{2}{3}\frac{\Sigma_{B}(q)}{q_s^2+q_3^2+\Sigma_{B}(q)}\right\}\bigg]\nonumber\\
&+& S[\L]
\eea
where
\bea
S[\L]=-\int d^3x N^{ab} \cf_{-+}^a\cf_{-3}^b -\tr\log\left(\frac{k}{4\pi}\delta^{ab}\partial_- - f^{ab}{}_c \Lambda_-^c\right)
\eea
Which is identical to \eqref{eq:sl}. Again the $S[\L]$ does not contribute to the gap equations at leading order in $N$ in the large-$N$ limit. The first line in Eq. \ref{Eq:BosonsFinalDual}, that is, the free energy of the scalar  plus Chern-Simons theory has been discussed in the literature extensively \cite{Aharony:2011jz,Aharony:2012ns,Jain:2012qi,Jain:2013gza}.

%%%%%%%%%%%%%%%%%%%%%%%%%%%%%%%%%%%%%%%%%%%%%%%%%%%%%%%%%%%%%%%
\section{Comments on level/rank duality with matter}\label{Sec:Comparison}
%%%%%%%%%%%%%%%%%%%%%%%%%%%%%%%%%%%%%%%%%%%%%%%%%%%%

Let us briefly recall in a schematic way the two dual theories

\bea
Z^F[\Lambda]&=&\int D\bar{\psi}D\psi \det N^{ab}\exp\bigg[\int d^3x \left(\bar{\psi}\slashed{\partial}\psi +V(\bar{\psi}\psi) + A_3^a(J_3^a+\cf_{+-}^a)\right)\bigg] \\
&=&\int D\bar{\psi}D\psi \det N^{ab}\exp\bigg[\int d^3x \left(\bar{\psi}\slashed{\partial}\psi +V(\bar{\psi}\psi) + N^{ab}(J_-^b+\cf_{-3}^b)(J_3^a+\cf_{+-}^a)\right)\bigg] \nn
\eea

\bea
Z^B[\Lambda]&=&\int D\bar{\phi}D\phi \det N^{ab}\exp\bigg[\int d^3x \left(-\bar{\phi}\nabla^2\phi +V(\bar{\psi}\psi) + A_3^a(J_3^a+\cf_{+-}^a)+A_3^a A_3^b\bar{\phi}T^aT^b\phi\right)\bigg] \nn\\
&=&\int D\bar{\phi}D\phi \det N^{ab}\exp\bigg[\int d^3x \left(-\bar{\phi}\nabla^2\phi +V(\bar{\psi}\psi) + N^{ab}(J_-^b+\cf_{-3}^b)(J_3^a+\cf_{+-}^a)\right. \nn\\
&+&\left. N^{ac}N^{bd}(J_-^c+\cf_{-3}^c)(J_-^d+\cf_{-3}^d)\bar{\phi}T^aT^b\phi\right)\bigg]
\eea

Since the path integral above can not be performed rigorously we now consider various approximations leading to or enriching the intuition of level/rank duality

%%%%%%%%%%%%%%%%%%%%%%%%%%%%%%%%%%%%%%%%%%%%%%%%%%%%%%%%%%%%%%%
\subsection{'t Hooft limit}
In this subsection we consider the 't Hooft limit: $N,k\to \infty$ with $\lambda =N/k$ held fixed. In the previous sections we have discussed the dual theory in the 't Hooft limit and we have found that it simplifies to the duality that has been discussed before plus an effective theory for the field $\Lambda_\mu^a$. The level/rank duality with matter is simply a consequence of the part independent of the field $\Lambda_\mu^a$ which has been discussed extensively in the literature plus, the form in which the tensor $N^{ab}$ enters in the action for $\Lambda_\mu^a$ . Namely, given that the level of the Chern-Simons action enters, in the large-$N$ limit as $N^{ab}\sim \delta^{ab} (\partial_-)^{-1}/k+...$ we conclude that level/rank, at this level simply exchanges $k\leftrightarrow N$.

%%%%%%%%%%%%%%%%%%%%%%%%%%%%%%%%%%%%%%%%%%%%%%%%%%%%%%%%%%%%%%%
\subsection{Small field approximation}

Level/rank duality with matter translate to a relation between the theories $Z^F[\Lambda]$ and $Z^B[\Lambda]$.  In the large-$N$ limit we simply reduce the duality to the statements present in the literature. We now discuss a small field expansion in $\Lambda$; this is, {\it a priori}, different from the large $N$ limit approximation. Since there is a lot of evidence about this duality in the case of zero Lagrange multiplier field $\Lambda=0$, it makes sense to explore a perturbative, in $\Lambda$, approach. If we formally take $\Lambda\to 0$ in the above expressions we simply recover the

\be
N^{ab}=\left(\frac{k}{2\pi}\delta^{ab}\partial_-\right)^{-1}.
\ee
Consequently, its determinant becomes a field-independent ($\Lambda$-independent) factor which we drop.
\bea
\label{Eq:ZF_L0}
Z^F[\Lambda\to 0]&=&\int D\bar{\psi}D\psi \exp\bigg[\int d^3p \left(\bar{\psi}\slashed{p}\psi +V(\bar{\psi}\psi)\right.\nn\\
&&+ \left.\frac{2\pi}{k}\frac{1}{p_-}J_-^a J_3^a\right)\bigg]
\eea

\bea
\label{Eq:ZB_L0}
Z^B[\Lambda\to 0]&=&\int D\bar{\phi}D\phi \exp\bigg[\int d^3p \left(-\bar{\phi}p^2\phi +V(\bar{\phi}\phi) \right. \nn\\
&&\left.+\frac{2\pi}{k}\frac{1}{p_-}J_-^a J_3^a +\left(\frac{2\pi}{k}\right)^2\frac{1}{p_-}\frac{1}{(p-q)_-} J_-^aJ_-^b\bar{\phi}T^aT^b\phi\right)\bigg]
\eea

These vector models have been extensively studied and the level/rank duality with matter has been verified in observables following from these models including the finite temperature free energies \cite{Giombi:2011kc,Aharony:2012ns,Jain:2013gza}. It is worth highlighting a key conceptual novelty of the actions in Eqs. (\ref{Eq:ZF_L0}) and \ref{Eq:ZB_L0}. Namely, their mild non-locality as witnessed by the powers of inverse momenta: $1/p_-$. This is similar to the nonlocality that appears in the dual theories where one schematically has, at leading order,  $\cf_{-3}^a \cf_{+-}^a/p_-$.

The momentum space equation for $A_3^a(p)$ takes a similar form in the fermionic Eq. \ref{Eq:A-Fermions} or bosonic case Eq. \ref{Eq:A-Bosons}:
\be
\frac{k}{2\pi}\delta^{ab}\, ip_- \, A_3^b(p) -f^{ab}{}_c \int \frac{d^3q}{(2\pi)^3}\L_-(q)^c \, A_3^b(p-q)+ J_-^a(p) -\cf_{-3}^a(p)=0.
\ee

In appendix \ref{Sec:MomentumSpace}, we solve this equation as an expansion in  $\Lambda$. The main result is that before taking the large $N$ limit there is a hierarchy pointing to an equivalence between the fermionic and bosonic theory. We do not pursue this analysis in this manuscript but it lays the foundation for a possible comparison at finite $N$ between the dual effective actions $Z^F[\Lambda]$ and $Z^B[\Lambda]$.
%%%%%%%%%%%%%%%%%
%1/N Corrections

%%%%%%%%%%%%%%%%%%%%%%%%%%%%%%%%55
\section{Comments on $\frac{1}{N} $ corrections}\label{Sec:BeyondN}
The discussion of $\frac{1}{N}$ corrections for the case $\L^a(x)=0$ is the standard one .i.e. one first integrates out the matter and gets an effective action in terms of the auxiliary  fields $\Sigma,\alpha $ \cite{Jain:2012qi}. The saddle points are computed from this effective action and finally  the effective action for the auxiliary fields is evaluated around the saddle points. By discarding the terms in this expansion which are higher order in $\frac{1}{N}$ one gets the non-local kinetic term for the auxiliary fields. The propagator for the auxiliary field is then used to perform the full path integral upto order $\frac{1}{N}$.\\
For our framework where $\L^a_\m(x)\ne0$, we do not want to integrate out everything, rather we want an effective action in terms of the dual field $\L^a_\m(x)$. This effective action can be expanded diagrammatically with  $\L^a_\m(x)$  vertices generating  external  lines for various $n-$point amplitudes. These amplitudes will cary group theory factors which will generate their $N$-dependence. It is in this sense that expansion of the effective action in powers of $\L^a_\m(x)$ can be considered a $\frac{1}{N^{\lambda}}$ expansion with $\lambda=0,1,...$.

As a first step towards computing $\frac{1}{N}$ corrections, the large-N saddle points are computed for $\mu,\alpha$'s from equations (\ref{Eq:saddlepsi}),(\ref{Eq:Sigmaphi}). In the next step the integrand of the original  path integral is expanded in power series in $\frac{1}{N}$  around the large-N saddle points.
%%%%%%%%%%%%%%%%%
%For small $\L$ approximation  equation (\ref{Eq:N1}) gives
%\bea
%N_1^e =\frac{2\pi}{k\partial_-}\mathcal{N}_0  \L_-^e,
%\eea
%then it is easy to see that $\mathcal{N}_0$ becomes
%\bea
%\mathcal{N}_0=\frac{2\pi}{k\partial_-}(1+\frac{1}{N}(\frac{2\pi}{k\partial_-})^2\L_-^a\L_-^a)^{-1}
%\eea
%and consequently
%\bea
%\mathcal{N}_1^a&=&-i (\frac{2\pi}{k\partial_-})^2\L_-^a(1+\frac{1}{N}(\frac{2\pi}{k\partial_-})^2\L_-^b\L_-^b)^{-1}\nonumber\\
%\mathcal{N}_2^a&=& (\frac{2\pi}{k\partial_-})^3\L_-^a(1+\frac{1}{N}(\frac{2\pi}{k\partial_-})^2\L_-^d\L_-^d)^{-1}d^{abc}\L_-^b\L_-^c
%\eea
%It is clear from the $k,N$ dependence of $\mathcal{N}_0 ,\mathcal{N}_1^a,\mathcal{N}_2^a$ that in the leading order and the next to leading order ($\frac{1}{N}$) contributions of large N expansion, with $\frac{N}{k}$ fixed, it is only $\mathcal{N}_0$ which will contribute.
\subsection{Bosonic effective action}
We start by considering the bosonic case
\bea
\label{Eq:Lphi}
\mathcal{L}_{\phi}&=&\int \frac{d^3p}{(2\pi)^3}\bigg[\bar{\phi}(-p)(p_s^2+p_3^2)\phi(p)+\frac{2\pi i}{k p_-}(j_-^a(p)\cf_{+-}^a(-p)+j_3^a(-p)\cf_{-3}^a(p))\nonumber\\
&+&\frac{2\pi i}{k p_-}(\delta^{ab}j_-^a(p)j_3^b(-p)+N^{ab}\cf_{-3}^a(p)\cf_{+-}^b(-p))
+\int \frac{d^3q}{(2\pi)^3}\int \frac{d^3r}{(2\pi)^3}(\frac{(2\pi i)^2}{k^2 p_- q_-}j_-^{a}(p)j_-^{b}(q)\nonumber\\
&+&\frac{(2\pi i)^2}{k^2 p_- q_-}\cf_{-3}^a(p)\cf_{-3}^b(q))\bar{\phi}(r)T^aT^b\phi(-p-q-r)\bigg]\nonumber\\
\eea
The above action explicitly contains $\frac{1}{N}$ corrections which are made manifest by also introducing $\alpha$ and 
$\Sigma$ through Hubbard-Stratonovich transformation by the identities (\ref{Eq:hubbard}) (\ref{Eq:hubbard2}).
Writing out everything explicitly upto $\frac{1}{N}$ corrections
\bea\label{Equation:Sphifull}
\mathcal{S}_{\phi}&=&\int \frac{d^3P_1}{(2\pi)^3}\bigg[\bar{\phi}(-P_1)((P_1)_s^2+(P_1)_3^2)\phi(P_1)\nonumber\\
&+&\frac{2\pi\lambda i}{N (P_1)_-}\int \frac{d^3P_2}{(2\pi)^3}(-\frac{1}{2}(2P_2+P_1)_-\bar{\phi}(-P_2)\cf_{+-}(-P_1)\phi(P_2+P_1)\nonumber\\
&+&(2P_2+P_1)_3\bar{\phi}(-P_2)\cf_{-3}(P_1)\phi(P_2-P_1))\bigg]\nonumber\\
&+&N \pi i \lambda\int \frac{d^3P_1}{(2\pi)^3} \frac{d^3q_1}{(2\pi)^3}\frac{d^3q_2}{(2\pi)^3}(\frac{(-P_1 + q_1 + q_2)_- (P_1 + q_1 + q_2)_3}{(-q_1 + q_2)_-}-\frac{1}{N}\frac{4(q_{1})_-(q_{2})_3}{(P_1)_-})\alpha(P,q_1)\alpha(-P,q2)\nonumber\\
&+&\int\frac{d^3P_1}{(2\pi)^3}N^{ab}(P_1)\cf_{-3}^a(P_1)\cf_{+-}^b(-P_1)\nonumber\\
&+&N(2\pi i )^2\lambda^2 \int\frac{d^3P_1}{(2\pi)^3}\frac{d^3P_2}{(2\pi)^3}\frac{d^3q_1}{(2\pi)^3}\frac{d^3q_2}{(2\pi)^3}\frac{d^3q_3}{(2\pi)^3}\bigg(\frac{(P_1 - P_2 - 2 (q_1 + q_2))_- (2 P_1 + P_2 + 2 (q_1 + q_3))_-}{(P_1 + P_2 - 2 q_1 + 2 q_2)_- (P_2 - 2 q_1 + 2 q_3)_-}\nonumber\\
&-&\frac{1}{N}\frac{-2 (q_1)_- (P_1 + 2 (P_2 + q_2 + q_3))_-}{(P_1)_- (P_1 - 2 q_2 + 2 q_3)_-}
-\frac{1}{N}\frac{2 (P_1 - P_2 - 2 (q_1 + q_2))_- (q_3)_-}{(P_1 + P_2)_- (P_1 + P_2 - 2 q_1 + 2 q_2)_-}\bigg)\nonumber\\
&\times&\alpha(P_1,q_1)\alpha(P_2,q_2)\alpha(-P_1-P_2,q_3)\nonumber\\
&+&N(2\pi i)^2\lambda^2\int\frac{d^3P_1}{(2\pi)^3}\frac{d^3P_2}{(2\pi)^3}\frac{d^3q_1}{(2\pi)^3}\frac{1}{ (P_1)_- (-P_1-P_2)_-}\frac{\alpha(P_2,q_1)}{N} \frac{\cf_{-3}^a(P_1)\cf_{-3}^a(-P_1-P_2)}{N^2}
%&+&\int \frac{d^3q}{(2\pi)^3}\int \frac{d^3r}{(2\pi)^3}(\frac{(2\pi i)^2}{k^2 p_- q_-}j_-^{a}(p)j_-^{b}(q)+\frac{(2\pi i)^2}{k^2 p_- q_-}\cf_{-3}^a(p)\cf_{-3}^b(q))\bar{\phi}(r)T^aT^b\phi(-p-q-r)\nonumber\\
\eea 
%where the composite variables $\alpha$ are functions of appropriate momenta
%Notice the following
%\bea
%&\frac{(2\pi i)^2}{k^2 p_- q_-}\cf_{-3}^a(p)\cf_{-3}^b(q))\bar{\phi}(r)T^aT^b\phi(-p-q-r)\nonumber\\
 %&=\frac{(2\pi i)^2}{k^2 p_- q_-}(p_-\L_3^a(p)-p_3\L_-^a(p))(q_-\L_3^a(q)-q_3\L_-^a(q))\bar{\phi}(r)T^aT^b\phi(-p-q-r)\nonumber\\
% &=\frac{(2\pi i)^2}{k^2}\L_3^a(p)\L_3^a(q)-\frac{(2\pi i)^2}{k^2}(\frac{q_3}{q_-}\L_3^a(p)\L_-^a(q)+\frac{p_3}{p_-}\L_3^a(q)\L_-^a(p))+\frac{(2\pi i)^2}{k^2}\L_-^a(p)\L_-^a(q)\frac{p_3q_3}{p_-q_-}\nonumber\\
%\eea
%Now the term $\frac{(2\pi i )^2}{k^2}\L_3^a(p)\L_3^a(q)$ seems to be a mass term for the scalar field $\phi$ and thus effectively making the theory non-conformal

\subsection{Fermionic effective action}
The fermonic starting point is
\bea
\label{Eq:Lpsi}
\mathcal{S}_{\psi}&=&\int \frac{d^3p}{(2\pi)^3}\bar{\psi}(-p)\gamma^{\mu}q_{\mu}\psi(p)+\frac{2\pi i}{k p_-}(j_-^a(p)\cf_{+-}^a(-p)+j_3^a(-p)\cf_{-3}^a(p))\nonumber\\
&+&\frac{2\pi i}{k p_-}\delta^{ab}(j_-^a(p)j_3^b(-p)+\cf_{-3}^a(p)\cf_{+-}^b(-p))\nonumber\\
\eea
It is crucial to highlight that in terms of currents we already see a difference with respect to the bosonic case, namely the $3rd$ line in equation (\ref{Eq:Lphi}).
Using the auxiliary field $\alpha$ for the fermonic case gives  upto $\frac{1}{N}$ corrections
\bea\label{Equation:Spsifull}
\mathcal{S}_{\psi}&=&\int \frac{d^3P_1}{(2\pi)^3}\bigg[\bar{\psi}(-P_1)\gamma^{\mu}(P_1)_{\mu}\psi(P_1)\nonumber\\
&+&\frac{2\pi\lambda }{N (P_1)_-}\int \frac{d^3P_2}{(2\pi)^3}(\bar{\psi}(-P_2)\cf_{+-}(-p)\gamma_-\psi(P_2+P_1)
+\bar{\psi}(-P_2)\cf_{-3}\gamma_3(P_1)\psi(P_2-P_1))\nonumber\\
&+&N 2 \pi i \lambda\int \frac{d^3q_1}{(2\pi)^3}\frac{d^3q_2}{(2\pi)^3}\bigg(\frac{1}{(q_1-q_2)_-}\alpha_-(P_1,q_1)\alpha(-P_1,q_2)-\frac{1}{N}\frac{1}{(P_1)_-}\alpha_-(P_1,q_1)\alpha_3(P_1,q_2)\bigg)\bigg]\nonumber\\
&+&\int \frac{d^3P_1}{(2\pi)^3}N^{ab}(P_1)\cf_{-3}^a(P_1)\cf_{+-}^b(-P_1)\nonumber\\
\eea
%Here $\alpha_-,\alpha_3$ are functions of momenta
%%%%%%%%%%%%%%%%%%%%%%%%%%%%%%%%%%%%%%%%%%%%%%%%%%%%%%%%%%%%%%%

\subsection{Saddle points}
Our strategy consists in constructing the effective actions as an expansion around the large$-N$ saddle points.
\bea\label{Eq:saddle}
&\mu(q)^{\phi}=0,\quad \alpha(q)=\frac{1}{q^2}\nonumber\\
&\mu(q)^{\psi}=\lambda \sqrt{2q_+q_-},\quad \mu_+(q)^{\psi}=-i \lambda^2  q_+\nonumber\\
&\alpha_-(q)^{\psi}=-\tr\frac{\gamma_-}{i \gamma_\m q^\m-i \lambda^2 q_+},\quad \alpha(q)^{\psi}=-\tr\frac{1}{i \gamma_\m q^\m+ \lambda \sqrt{2q_+q_-}}
\eea
and
 the propagator for bosonic theory after including large$-N$ corrections
\bea
\alpha_{B}(p)=\frac{1}{p^2}
\eea
For the  fermonic  propagator after including large N corrections
\bea\label{Equation:psipropagator}
\alpha_{F}(p)=\frac{-i \bigg(\gamma_- (p)_+(1-\lambda^2)+(p)_-\gamma_++(p)_3\gamma_3\bigg)+\lambda(2 p_+p_-)^{\frac{1}{2}}}{p^2}
\eea

\subsection{1-loop contribution to subleading order in $\frac{1}{N}$}\label{alphasigma}
Reminding ourselves of the gap equations for fermionic theory
\bea
&\Sigma_{F,+}(p)=-2 \pi i \lambda \int d^3 x \frac{1}{\partial_-}\tr(\frac{1}{\gamma^{\mu}\partial_{\mu}+\Sigma_{F}+\frac{2\pi}{k\partial_-}(\cf_{+-}^bT^b\gamma_-+\cf_{-3}^bT^b\gamma_3)}),\nonumber\\
&\Sigma_{F,I}(p)=2 \pi i \lambda \int d^3 x \frac{1}{\partial_-}\tr(\frac{\gamma_-}{\gamma^{\mu}\partial_{\mu}+\Sigma_{F}+\frac{2\pi}{k\partial_-}(\cf_{+-}^bT^b\gamma_-+\cf_{-3}^bT^b\gamma_3)}),
\eea
with $\Sigma_F=\Sigma_+\gamma_-+\Sigma_0$
and for bosonic theory
\bea
\Sigma_B(p)&=&\int\frac{d^3 q_1}{(2\pi)^3}\frac{d^3 q_2}{(2\pi)^3}\bigg[ C_2(p,q_1,q_2)+C_2(q_1,p,q_2)+C_2(q_1,q_2,p) \bigg]\alpha_B(q_1)\alpha_B(q_2),\nonumber\\
& & 
\eea
where in position space $\alpha_B$ can be written as
\bea
\alpha_B=\tr\frac{1}{\partial^2+\Sigma_{B}+\frac{2\pi}{k\partial_-}(\cf_{+-}^bT^b\gamma_-+\cf_{-3}^bT^b\gamma_3)+\frac{(2\pi)^2}{(k\partial_-)^2}\cf_{-3}^a\cf_{-3}^bT^aT^b},
\eea
where now the $\tr$ is a trace on both the position space and color space.\\
Next making the shifts around the saddle points (\ref{Eq:saddle})
\bea
\Sigma_+\to \Sigma_+-i\lambda^2 q_+,\qquad \Sigma_0\to \Sigma_0+\lambda q_s,
\eea
where now $\Sigma_+,\Sigma_0$ are of order $\mathcal{O}(\frac{1}{N})$
\bea
S_{eff}^{\psi}&=&N V\bigg(- \int d^3x\log\det(\gamma^{\mu}\partial_{\mu}-i \lambda^2 \partial_+\gamma_-+\lambda \partial_s-\frac{2\pi}{k\partial_-}(\cf_{+-}^bT^b\gamma_-+\cf_{-3}^bT^b\gamma_3)+\gamma_{-} \Sigma_{+}
+\Sigma_{0})
\nonumber\\
&+&\int \frac{d^3p_1}{(2\pi)^3}\frac{d^3p_2}{(2\pi)^3}\frac{8\pi i N}{k (p_1-p_2)_-}\alpha_-(p_1)\alpha_0(p_2) )
\nonumber\\
&+&\int\frac{d^3q}{(2\pi)^3}( - 2 (\Sigma_+(q)-i\lambda^2 q_+)\alpha_-(q)- 2 (\Sigma_0(q)+\lambda q_s)\alpha_0(q))
\bigg)
\eea
So the fermion effective action evaluated at the large$-N$ $(N\to\infty)$  saddle points becomes
\bea\label{Equation:Spsisaddle}
S_{eff}^{\psi}|_{saddle}&=&N V\bigg(- \int d^3x\log\det(\gamma^{\mu}\partial_{\mu}-i \lambda^2 \partial_+\gamma_-+\lambda \partial_s-\frac{2\pi}{k\partial_-}(\cf_{+-}^bT^b\gamma_-+\cf_{-3}^bT^b\gamma_3))
\nonumber\\
&+&\int \frac{d^3p_1}{(2\pi)^3}\frac{d^3p_2}{(2\pi)^3}\frac{8\pi i N}{k (p_1-p_2)_-}\alpha_-(p_1)\alpha_0(p_2) )
\nonumber\\
&+&\int\frac{d^3q}{(2\pi)^3}(  2 i\lambda^2 q_+\alpha_-(q)- 2 \lambda q_s \alpha_0(q))
\bigg).
\eea
Similarly, expanding around the large N saddle point of bosonic auxiliary field $\Sigma_B$
\bea
S_{eff}^{\phi}&=& N V\bigg[\int d^3 x \tr\log (\partial_s^2+\partial_3^2+\Sigma_{B}+\frac{2\pi}{k\partial_-}(\cf_{+-}^bT^b\gamma_-+\cf_{-3}^bT^b\gamma_3)+\frac{(2\pi)^2}{(k\partial_-)^2}\cf_{-3}^a\cf_{-3}^bT^aT^b)\nonumber\\
&+&\int\frac{d^3q_1}{(2\pi)^3}\frac{d^3q_2}{(2\pi)^3}\frac{d^3q_3}{(2\pi)^3}C_2(q_1,q_2,q_3)\alpha_B(q_1)\alpha_B(q_2)\alpha_B(q_3)-\int\frac{d^3q}{(2\pi)^3}\Sigma_B(q)\alpha_B(q)
\bigg],\nonumber\\
\eea
where $\Sigma_B$ is of order $\mathcal{O}(\frac{1}{N})$.
Now at the  large N $(N\to\infty)$ saddle point $\Sigma_B=0$
\bea
S_{eff}^{\phi}|_{saddle}&=& N V\bigg[\int d^3 x \tr\log (\partial_s^2+\partial_3^2+\frac{2\pi}{k\partial_-}(\cf_{+-}^bT^b\gamma_-+\cf_{-3}^bT^b\gamma_3)+\frac{(2\pi)^2}{(k\partial_-)^2}\cf_{-3}^a\cf_{-3}^bT^aT^b)\nonumber\\
&+&\int\frac{d^3q_1}{(2\pi)^3}\frac{d^3q_2}{(2\pi)^3}\frac{d^3q_3}{(2\pi)^3}C_2(q_1,q_2,q_3)\alpha_B(q_1)\alpha_B(q_2)\alpha_B(q_3)
\bigg].\nonumber\\
\eea

%IMPORTANT!\\
%The appearance of the dual field strength $\cf_{\m\n}^a=\partial_\m\L_\n^a-\partial_\n\L_\m^a$ in the $S^{\psi}_{eff},S^{\phi}_{eff}$ is peculiar. Invoking the fact that the saddle point value of $\L_\n^a$ should be Poincare invariant implies that 
%\bea
%\L_\m^a=0\quad at\quad the \quad saddle\quad point
%\eea
%This has the important consequence that 
\subsection{A small digression to Functional Determinants}
The functional determinants  arising in QFT calculations can be expanded in terms of Feynman diagrams.
For example the differential operator for a massive Dirac field in the background of auxiliary field $\phi(x)$ can be written as
\bea
Z(\phi)=Det(-i \partial_\m\gamma^{\m}+m-g\phi(x))=e^{-\sum_{n=1}^{\infty}\frac{1}{n}\tr G^n}
\eea
upto a factor which is independent of the background field $\phi(x)$. Where
\bea
\tr G^n=g^n\int d^3x_1d^3x_2...d^3x_n tr(S(x_1-x_2)\phi(x_2)...S(x_n-x_1)\phi(x_1))
\eea
where $S$ is the free field Dirac propagator $S(p)=\frac{1}{i\gamma_\m p^{\m}+m}$.\\
This is precisely the diagrammatic approach we develop in appendix (\ref{Sec:diagrammatics}) as adapted to our case.
\section{Evaluation of the Effective actions at the saddle points}\label{Sec:EA}
It is obvious from the  saddle point solutions (\ref{Eq:saddle}) that the fermonic saddle point depends on the large $N$ 't Hooft coupling constant $\lambda=\frac{N}{k}$, whereas the bosonic saddle point is trivial. As a consequence when the fermonic effective  action $S^{\psi}_{eff}$ is evaluated at the saddle point and one tries to get the effective interactions for the dual field $\L$ by evaluating $1-point$, $2-point$...$n-point$ functions, certain extra interactions are generated which depend on the  various powers of the coupling constant $\lambda$. However, it can be shown that these extra Feynman integrals vanish. 
\\
To elaborate let us consider $S^{\psi}_{eff}$ 
\bea\label{Eq:fermi}
S_{eff}^{\psi}|_{saddle}&=&N V\bigg(- \int d^3x\tr\log(\gamma^{\mu}\partial_{\mu}-i \lambda^2 \partial_+\gamma_-+\lambda \partial_s)\nonumber\\
&+&\tr\log(1-\frac{1}{(\gamma^{\mu}\partial_{\mu}-i \lambda^2 \partial_+\gamma_-+\lambda \partial_s)}\frac{2\pi}{k\partial_-}(\cf_{+-}^bT^b\gamma_-+\cf_{-3}^bT^b\gamma_3))
\nonumber\\
&+&\int \frac{d^3p_1}{(2\pi)^3}\frac{d^3p_2}{(2\pi)^3}\frac{8\pi i N}{k (p_1-p_2)_-}\alpha_-(p_1)\alpha_0(p_2) )
\nonumber\\
&+&\int\frac{d^3q}{(2\pi)^3}(  2 i\lambda^2 q_+\alpha_-(q)- 2 \lambda q_s \alpha_0(q))+\mathcal{O}(\frac{1}{N})\quad terms
\bigg).
\eea
Now, the second line in eq. (\ref{Eq:fermi}) can be evaluated by Feynman diagram technique to find $n-point$ function of the dual field $\L$ and obviously these diagrams will contain $\lambda$ dependence arising from the full fermonic propagator. However, the claim made above is that at least to order $\mathcal{O}(\frac{1}{N})$ in the diagrammatic expansion this fermonic propagator can be replaced by the free propagator in the second line.
\bea\label{Eq:fermi2}
S_{eff}^{\psi}|_{saddle}&=&N V\bigg(- \int d^3x\tr\log(\gamma^{\mu}\partial_{\mu}-i \lambda^2 \partial_+\gamma_-+\lambda \partial_s)\nonumber\\
&+&\tr\log(1-\frac{1}{(\gamma^{\mu}\partial_{\mu})}\frac{2\pi}{k\partial_-}(\cf_{+-}^bT^b\gamma_-+\cf_{-3}^bT^b\gamma_3))
\nonumber\\
&+&\int \frac{d^3p_1}{(2\pi)^3}\frac{d^3p_2}{(2\pi)^3}\frac{8\pi i N}{k (p_1-p_2)_-}\alpha_-(p_1)\alpha_0(p_2) )
\nonumber\\
&+&\int\frac{d^3q}{(2\pi)^3}(  2 i\lambda^2 q_+\alpha_-(q)- 2 \lambda q_s \alpha_0(q))+\mathcal{O}(\frac{1}{N})\quad terms
\bigg)
\eea
Now we write the bosonic action to make a comparison
\bea
S_{eff}^{\phi}|_{saddle}&=& N V\bigg[\int d^3 x \tr\log (\partial_s^2+\partial_3^2)\nonumber\\
&+&\tr\log(1+\frac{1}{\partial_s^2+\partial_3^2}(\frac{2\pi}{k\partial_-}(\cf_{+-}^bT^b\gamma_-+\cf_{-3}^bT^b\gamma_3)+\frac{(2\pi)^2}{(k\partial_-)^2}\cf_{-3}^a\cf_{-3}^bT^aT^b))\nonumber\\
&+&\int\frac{d^3q_1}{(2\pi)^3}\frac{d^3q_2}{(2\pi)^3}\frac{d^3q_3}{(2\pi)^3}C_2(q_1,q_2,q_3)\alpha_B(q_1)\alpha_B(q_2)\alpha_B(q_3)+\mathcal{O}(\frac{1}{N})\quad terms
\bigg]\nonumber\\
\eea
It is clear that the two effective actions for $\L$ differ in the term $\frac{(2\pi)^2}{(k\partial_-)^2}\cf_{-3}^a\cf_{-3}^bT^aT^b$
This term to leading order in large N generates a tadpole diagram which gives zero contribution in the Dimensional regularization scheme which we use. However, from next to leading order in the large $N$ expansion this vertex generates Feynman diagrams which have no counter part in the fermonic theory. Therefore the  theories start deviating from duality  at this order. \\
There is an important  and rather crucial point here. When the  vertex $\frac{(2\pi)^2}{(k\partial_-)^2}\cf_{-3}^a\cf_{-3}^bT^aT^b$ is not present, the set of diagrams generated might match. One could argue that this is indeed what happens for the critical theory because the origin of this vertex is the same as that of $(\bar{\phi}\phi)^2$ coupling which arises from integrating out  the Chern Simons gauge field $A_\m^a(x)$ in the bosonic action.  It will be interesting to analyze this issue in more detail but it goes beyond our interests.

\section{Conclusions}\label{Sec:Conclusions}
%%%%%%%%%%%%%%%%%%%%%%%%%%%%%%%%%%%%%%%%%%%%%%%%%%%%

In this manuscript we have used a constructive approach to  investigate connections among 3D vector theories. We have, in particular, addressed some aspects of bosonization in three dimensions and the level/rank duality. %the role of coupling different matter. 
We have considered integration in various limits but most prominently the large $N$ limit employing some of the standard techniques. Besides defining formally the dual theories to the fermionic and scalar $U(N)$ vector models we were able to identify the two dual theories and show that they agree in the large $N$ limit exhibiting explicitly a  level/rank duality for these systems. This provides a concrete evidence for the validity of the conjectured duality among the different vector models in the large $N$ limit.

There are a number of questions that would be interesting to pursue. The large $N$ limit always raises the question of $1/N$ corrections. We have shown that the large $N$ limit is well defined. A deeper study of $1/N$ corrections, for which our work is a first step,  is particularly pressing given the claims that the level/rank duality with matter might extend to finite $N$  \cite{Aharony:2012nh,Jain:2013gza}.  In sections \ref{Sec:BeyondN} and \ref{Sec:EA} we present an admittedly scant approach to the effective actions beyond the leading order in $N$. We find that, generically, the effective fermionic and bosonic actions do differ at this level. The main culprit being the contribution of a particular vertex in the diagrammatic approach as elaborated in appendix (\ref{Sec:diagrammatics}). We have provided a scant evidence that the duality fails to hold beyond the large $N$ leading order. 

Given that one of the strongest evidence for level/rank duality with matter consists in the matching of the corresponding finite temperature free energies \cite{Giombi:2011kc,Aharony:2012ns,Jain:2013gza}, it would be interesting to consider the path integral dualization procedure used here for finite temperature backgrounds. It was pointed almost two decades ago \cite{Rocek:1991ps, Alvarez:1993qi} that the dualization procedure in non-contractible spaces, where pure gauge  connections are not necessarily trivial, need a careful treatment. This pertains, in particular, for spaces with $\pi_1(X) \neq 0$ which is precisely the case with the thermal circle. We hope to return to this interesting question in the near future. 

Lastly, even though our approach is consistent with the vector model/higher spin duality, it does not directly address this duality.  One important question would be to identify the role of the  extra dimension in Vasiliev's theory under these dualities. There are some tantalizing hints connecting the non-local aspects of the duality with a higher dimensional theory. It would be very interesting to pursue this line. Although a more covariant approach, than the one presented here and rooted in the light-cone gauge, would arguably be required.

%%%%%%%%%%%%%%%%%%%%%%%%%%%%%%%%%%%%%%%%%%%%%%%%%%%%%%%%%%%%%%%
\section*{Acknowledgments}
%%%%%%%%%%%%%%%%%%%%%%%%%%%%
We are particularly thankful to Guillermo Silva for various piercing comments. We are grateful to Aki Hashimoto, Yolanda Lozano, Shiraz Minwalla, Jos\'e Francisco Morales, Jeff Murugan, Cheng Peng, Guilherme Pimentel and  Cobi Sonnenschein for comments.

\begin{appendix}

%%%%%%%%%%%%%%%%%%%%%%%%%%%%%%%%%%%%%%%%%%%%%%%%%%%%%%%%%%%%%%%
\section{Integrating fermions and Chern-Simons theory}
%%%%%%%%%%%%%%%%%%%%%%%%%%%%%%%%%%%%%%%%%%%%%%%%%%%%
The integral over fermions in 3d is a fairly standard computation, see for example \cite{Dunne:1998qy}. We reproduce part of it here just to render the manuscript more self-contained and to clarify our normalizations.
\bea
S_{eff}[A,m]&=&N_f \ln \det (i\slashed{\partial}+\slashed{A}+m) \nonumber \\
&=& N_f \ln \det\big[(i\slashed{\partial}+m)(1+(i\slashed{\partial}+m)^{-1}\slashed{A})\big]\nonumber \\
&=&N_f \ln\det(i\slashed{\partial}+m)+ N_f \log \det(1+(i\slashed{\partial}+m)^{-1}\slashed{A}) \nonumber \\
&\sim &N_f \ln \det(1+(i\slashed{\partial}+m)^{-1}\slashed{A}),
\eea
where the first term in the penultimate line can be neglected because it is $A$ independent and we are looking for an effective theory of $A$. Expanding the determinant: 
\bea
S_{eff}[A,m]&=&N_f \tr\ln(1+(i\slashed{\partial}+m)^{-1}\slashed{A}) \nonumber \\
&\approx& N_f \tr \left(\frac{1}{i\slashed{\partial}+m}\slashed{A}\right)
-\frac{1}{2}N_f \tr\left(\frac{1}{i\slashed{\partial}+m}\slashed{A}\frac{1}{i\slashed{\partial}+m}\slashed{A}\right)+\ldots
\eea

The first term corresponds to a tadpole, the second term contributes to
\bea
S_{eff}^{quadratic}[A,m]&=& \frac{1}{2}N_f \int\frac{d^3p}{(2\pi)^3}A_\mu(-p)\Gamma^{\mu\nu}(p,m)A_\nu(p), \nonumber\\
\Gamma^{\mu\nu}(p,m)&=&\int \frac{d^3k}{(2\pi)^3}\tr\gamma^\mu \frac{\slashed{p}+\slashed{k}-m}{(p+k)^2+m^2}\gamma^\nu \frac{\slashed{k}-m}{k^2+m^2}. \nonumber\\
\eea
The parity odd part can be computed exactly in  the large mass $(m\to \infty)$, long wavelength limit $(p\to 0)$.
\bea
\Gamma^{\mu\nu}(p,m) &\sim & \epsilon^{\mu\nu\rho}p_\rho  \frac{1}{4\pi}\frac{m}{|m|} +{\cal O}\left(\frac{p^2}{m^2}\right).
\eea

The three gluon diagram leads to
\bea
S^{cubic}_{eff}[A,m]&=& \frac{N_f}{3}\int \frac{d^3p_1 d^3 p_2}{(2\pi)^6}A_\mu(p_1) A_\nu(p_2)A_\rho(-p_1-p_2)\Gamma^{\mu\nu\rho} (p_1, p_2, m), \\
\Gamma^{\mu\nu\rho} (p_1, p_2, m)&=&\int \frac{d^3k}{(2\pi)^3}\tr\gamma^\mu \frac{\slashed{p}_1+\slashed{k}-m}{(p_1+k)^2+m^2}\gamma^\nu \frac{\slashed{p}_1+\slashed{p}_2+\slashed{k}-m}{(p_1+ p_2+k)^2+m^2}\gamma^\rho \frac{\slashed{k}-m}{k^2+m^2}, \nonumber
\eea

The relevant odd  component, in the large mass limit $(m\gg p_1, p_2)$:
\be
\G^{\mu\nu\rho}_{odd}(p_1, p_2,m) \sim -i\frac{1}{4\pi}\frac{m}{|m|}\epsilon^{\mu\nu\rho}+ {\cal O}\left(\frac{p^2}{m^2}\right).
\ee

%%%%%%%%%%%%%%%%%%%%%%%%%%%%%%%%%%%%%%%%%%%%%%%%%%%%%%%%%%%%%%%%%%%%%%%%%%%%%%%%
\section{Functional integration and an obstruction}
%%%%%%%%%%%%%%%%%%%%%%%%%%%%%%%%%%%%%%%%%%%%%%%%%%%%%%%%%%%%%%%%%%%%%%%%%%%%%%%%%%
Schematically the integrals that are being computed are often of the form:
\bea
&&\int DA_+^a DA_3^a \exp \left(A_+^a (f^{abc}\L_-^c A^{b}_3+J_-^a +\cf_{-3}^a)+ A_3^a (J_3^a+\cf_{+-}^a)\right) \nonumber \\
&=& \int DA_3^a \delta (f^{abc}\L_-^c A^{b}_3+J_-^a+\cf_{-3}^a )\exp \left( A_3^a (J_3^a+\cf_{+-}^a)\right) \nonumber \\
&=& \left(\det( f^{ab}{}_c \Lambda_-^c )\right)^{-1} \exp\left(M^{ab}(J_-^a +\cf_{-3}^a)(J_3^a+\cf_{+-}^a\right).
\eea
Note that the last integration over $A_3^a$ requires that we solve the argument of the delta function but then the results will contain $\det^{-1}( f^{ab}{}_c \Lambda_-^c )$.
Note that this contribution to the measure is reinterpreted in two-dimensions as a dilaton shift. Namely, 
\be
\Phi \mapsto \Phi -\frac{1}{2}\ln \det M^{ab}.
\ee

%%%%%%%%%%%%%%%%%%%%%%%%%%%%%%%%%%%%%%%%%%%%%%%%%%%%%%%%%%%
\subsection{Fermions: A singularity for light-cone integration without Chern-Simons}
%%%%%%%%%%%%%%%%%%%%%%%%%%%%%%%%%%%%%%%%%%%%%%%%%%%%%%%%%%%%%%%%
When treating gauge fields in three dimensions there are usually significant simplifications in the light-cone gauge; this can be seen in the Yang-Mills case \cite{Nair:2013hpa} but also in the Chern-Simons coupled to matter context that we are interested in \cite{Giombi:2011kc}. Here we consider a fermionic action without Chern-Simons coupling $(k\to 0$). Given the expected simplifications  it is tempting to attempt to first of all integrate the gauge field in the master action (\ref{Eq:FermionsMaster}).  In the light-cone gauge, $A_-=0$, the action takes the form ($k=0$):
\bea
\label{Eq:FermionLC}
S&=& \bar{\psi}\slashed{\partial}\psi +V(\bar{\psi}\psi) \nonumber \\
&+& A_+^a\left(J_-^a +\cf_{-3}^a +f_{abc}\,\L_-^bA_3^c\right) + A_3^a \left(J_3^a +\cf_{+-}^a\right),
\eea
where $J_\mu ^a = -i \bar{\psi}\gamma_\mu T^a\psi$ and $\cf_{\mu\nu}^a =\partial_\mu \L_\nu^a -\partial_\nu \L_\mu^a$. The resulting action is linear in $A_+$ and we could consider integrating it out as in the previous subsection which would yield

\be
\label{Eq:obstruction}
\cf_{-3}^a +J_-^a - f^a{}_{bc}A_3^b \L_-^c=0.
\ee
However, the matrix $f_{ab}=f_{abc}\Lambda_-^c$ is not invertible. One can check that its determinant vanishes\footnote{We thank Y. Lozano for important comments regarding gauge invariance on this point.}. Similar structures with the corresponding determinants appear in the context of non-Abelian T-duaity in string theory    \cite{delaOssa:1992vc,Alvarez:1993qi,Lozano:2011kb}.

The obstruction above, nevertheless, teaches us an important lesson about the intermediate structure of the dual theory. If we were able to solve Eq. (\ref{Eq:obstruction}) we would obtain that $A_3^a \sim J_-^a+\cf_{-3}^a$; substituting this back into the action would lead to four-fermion interactions. The singularity also teaches us that $(k\neq 0)$, that is,  coupling to a Chern-Simons  term precisely provides the regularization we seek that allows to invert Eq. (\ref{Eq:obstruction}) and that, indeed, eliminating the gauge field leads to an intermediate action containing four-fermion interactions arising as the product of fermionic currents.

%%%%%%%%%%%%%%%%%%%%%%%%%%%%%%%%%%%%%%%%%%%%%%%%%%%%%%%%%%%
\subsection{Bosons: A singularity for light-cone integration without Chern-Simons}

Let us show  that, as in the fermionic case, there is an obstruction to using the light-cone gauge and integrating the gauge field degrees of freedom in the absence of a Chern-Simons term $(k\to 0)$. One might have expected that perhaps the difference in couplings could have ameliorated this problem. In the light-cone gauge the action becomes ($k=0)$:
\bea
S&=& \int d^3x \bigg[-\bar{\phi}\left(2\partial_+\partial_-+\partial_3^2\right)\phi +A_+^aJ_-^a+A_3^a J_3^a + A_3^a A_3^b \bar{\phi}T^a T^b \phi \nonumber \\
&-&\partial_-A^a_+\L^a_3+\partial_- A^a_3 \L^a_+  + (\partial_3A_+^a -\partial_+A_3^a +f_{abc}A_3^b A_+^c)\L_-^a\bigg].
\eea
where $J_\mu^a$ was defined above in Eq. (\ref{Eq:BosonCurrent}).

The action is linear in $A_+^a$ whose equation of motion leads to
\be
\cf_{-3}^a +J_-^a - f^a{}_{bc}A_3^b \L_-^c=0.
\ee
This equation is similar to Eq. (\ref{Eq:obstruction}), the only difference is the construction the current which was fermionic in the case of Eq. (\ref{Eq:obstruction}) and it is bosonic here. The problem is, as before, that $f_{ab}=f_{abc}\L_-^c$ is not invertible. Note that the presence of quadratic in $A_3^a$ terms which are absent in the fermionic case do not modify the above statement as it relies only in integration over $A_+^a$.

We speculate that this singular limit, $k\to 0$, is teaching us about the scalings for which the theory can ultimately be well defined. It would be particularly interesting to return to this question in the face of claims that the level/rank duality with matter holds for finite $N$ \cite{Giombi:2011kc,Aharony:2012ns,Jain:2013gza}. Clearly, the scaling of $k$ becomes an important issue.

%%%%%%%%%%%%%%%%%%%%%%%%%%%%%%%%%%%%%%%%%%%%%%%%%%%%%%%%%%%%%%%%%%%%%%%%%%%%%%%%
\section{Momentum space actions}\label{Sec:MomentumSpace}
%%%%%%%%%%%%%%%%%%%%%%%%%%%%%%%%%%%%%%%%%%%%%%%%%%%%%%%%%%%%%%%%%%%%%%%%%%%%%%%%%%
We consider the presentation of various actions in momentum space. We use:
\be
\Phi(x)=\int \frac{d^3 p}{(2\pi)^3}e^{ipx}\Phi(p).
\ee

The action in momentum space following from a fundamental fermion coupled to $U(N)$ level $k$ Chern-Simons action in the light-cone gauge was discussed in \cite{Jain:2012qi} in a slightly more general setup of supersymmetric theories.
\be
S=\int d^3x \left(\bar{\psi}\slashed{\partial} \psi + A_3^a J_3^a\right),
\ee
where $A_3^a$ satisfies the equation
\be
\frac{k}{2\pi}\partial_- A_3^a+J_-^a=0.
\ee
In momentum space the above equation is solved by
\be
A_3^a(p)=\frac{2\pi i }{k\, p_-}J_-^a (p).
\ee
Plugging this expression in the above action, it can be rewritten as:
\bea
S&=& \int \frac{d^3 q}{(2\pi)^3} \bar{\psi}(-q) \slashed{q} \psi(q)  \nonumber \\
&+&{\rm Tr} \int \frac{d^3 p}{(2\pi)^3}\frac{d^3 r}{(2\pi)^3}\frac{d^3 s}{(2\pi)^3}\frac{-2\pi i }{k p_-}
\bar{\psi}(-r)\gamma_-T^a \psi (p+r) \bar{\psi}(-s)\gamma_3 T^a \psi(-p+s).
\eea

The full action that we consider in the main text is:
\bea
S&=& \int d^3 x \bigg[\frac{ik}{4\pi}\epsilon^{\mu\nu\rho}\tr\left(A_\mu \partial_\nu A_\rho -\frac{2i}{3}A_\mu A_\nu A_\rho\right) \nonumber \\
&+& \bar{\psi}\left(\slashed{\partial}- iA_\mu^a T^a\gamma^\mu  + m\right)\psi + V(\bar{\psi}\psi)+ \epsilon^{\mu\nu\rho}\tr (F_{\mu\nu}\L_\rho)\bigg].
\eea

The momentum space equation for $A_3^a(p)$ is
\be
\frac{k}{2\pi}\delta^{ab}\, ip_- \, A_3^b(p) -f^{ab}{}_c \int \frac{d^3q}{(2\pi)^3}\L_-(q)^c \, A_3^b(p-q)+ J_-^a(p) -\cf_{-3}^a(p)=0.
\ee

\subsection{ Perturbative solution in $\Lambda$:} 
Assuming that the zeroth oder solution, that is, the solution  without $\L$ is known: $A_3^a(p)=2\pi i J_-^a(p)/(k\, p_-)$ . Then, the Neumann  series solution  of the above integral equations  for $A_3^a(p)$,upto third order in $\frac{1}{k}$ (where $k$ is the Chern-Simons level), will be

\begin{align}
\label{Eq:A_L}
A_3(p)^a =& \frac{2\pi i}{k\, p_-}J_-^a(p) \nonumber \\
+&\frac{2\pi i}{k\, p_-}\cf_{3-}^a(p) -\frac{2\pi i}{k\, p_-}f^{a}{}_{bc}\int \frac{d^3 q }{(2\pi)^3}\L_-^c(q)\frac{2\pi i}{k\, (p-q)_-} J^b_-(p-q)
\nonumber \\
-&  \frac{2\pi i}{k\, p_-}f^{a}{}_{bc}\int \frac{d^3 q }{(2\pi)^3}\L_-^c(q)\frac{2\pi i}{k\, (p-q)_-}\cf_{3-}^b(p-q) )
\nonumber\\
+\frac{2\pi i}{k p_-}f^{aa_1a_2}f^{a_1a_3a_4}&\int\frac{d^3 p_1}{(2\pi)^3}\frac{d^3 p_2}{(2\pi)^3}\L_-^{a_2}(p_1)\L_-^{a_4}(p_2) \frac{2\pi i}{k\, (p-p_1)_-} \frac{2\pi i}{k\, (p-p_1-p_2)_-}J^{a_3}_-(p-p_1-p_2))\nonumber \\
+\frac{2\pi i}{k p_-}f^{aa_1a_2}f^{a_1a_3a_4}&\int\frac{d^3 p_1}{(2\pi)^3}\frac{d^3 p_2}{(2\pi)^3}\L_-^{a_2}(p_1)\L_-^{a_4}(p_2) \frac{2\pi i}{k\, (p-p_1)_-} \frac{2\pi i}{k\, (p-p_1-p_2)_-} \cf_{3-}^{a_3}(p-p_1-p_2) \nonumber \\
\end{align}
Note that in the above equation, the first line is the zeroth (no factors of $\Lambda$) order solution. The first line contains one power of $\Lambda$ while the second and third contain two powers of $\Lambda$.  The third line contains three powers of $\Lambda$; this way of organizing the expansion teaches us the general form in which powers of $\Lambda$ can appear, including when they appear only through $\cf_{3-}$.

A schematic form of the solution for $A_3$ is as follows:
\bea
\label{Eq:Apert}
A_3(p)&=& \frac{1}{kp_-}J_- \nn\\
&+& \frac{1}{k p_-}\cf_{3-}+ \frac{1}{k^2}f\int \Lambda_-J_- \nn\\
&+&\frac{1}{k^2} f\int \Lambda_-\cf_{3-}+\frac{1}{k^3}f^2\int (\Lambda_-)^2 J_- \nn\\
&+&\frac{1}{k^3} f^2\int (\Lambda_-)^2 \cf_{3-} +\frac{1}{k^4} f^3\int (\Lambda_-)^3J_-  \nn\\
&+& \ldots  + ... \nn\\
&+& \frac{1}{k^n} f^{n-1}\int (\Lambda_-)^{n-1}\cf_{3-}+ \frac{1}{k^{n+1}}f^n \int (\Lambda_-)^n J_-
\eea

After plugging the solution given by Eq. (\ref{Eq:A_L})  back into the action we will  analyze the terms of order $\frac{1}{k^2}$ and $\frac{1}{k^3}$   and show that they are subleasing in the Large-N limit in the singlet sector of the theory. The key identity we will use repeatedly  is 
\\
\bea
(T^a)^i_{ j} (T^b)^k_l=\frac{\delta^{ab}}{N^2-1}(\delta^i_l \delta^k_j-\frac{1}{N}\delta^i_j \delta^k_l)+\frac{N}{2(N^2-4)}d^{abc}(T^c)^i_{ l}\delta^k_j +
\nonumber\\
\frac{N}{2(N^2-4)}d^{abc}(T^c)^k_{ j}\delta^i_l -\frac{1}{N^2-4}d^{abc}(T^c)^i_{ j}\delta^k_l-\frac{1}{N^2-4}d^{abc}(T^c)^k_{ l}\delta^i_j
\nonumber\\
+\frac{i}{2N}f^{abc}T^c)^i_{ l}\delta^k_j-\frac{i}{2N}f^{abc}T^c)^k_{ j}\delta^i_l
\eea

Before writing out the result of plugging the expression for $A_3$ in  Eq. (\ref{Eq:A_L})  in the action let us first describe, schematically, the possible terms:

\bea
S&=& S^{(A)}+S^{(B)}\nn\\
&=&A_3^a J^a_3 + A_3 \bar{\phi}A_3\phi \nn\\
\eea
with
\bea
S^{(A)}&=&\frac{1}{k}J_- J_3 \nn\\
&+&\frac{1}{k}\cf_{3-}J_3 +\frac{1}{k^2}f \Lambda_- J_-J_3 \nn\\
&+& \frac{1}{k^2}f\Lambda_-\cf_{3-}J_3 + \frac{1}{k^3}f^2 (\Lambda_-)^2J_-J_3 \nn\\
&+& \ldots \nn\\
&+& \frac{1}{k^n}f^{n-1}(\Lambda_-)^{n-1}\cf_{3-}J_3 +\frac{1}{k^{n+1}}f^n \,\, (\Lambda_-)^n J_- J_3
\eea
This structure follows immediately from the structure of the solution of $A_3$ in Eq. (\ref{Eq:Apert}). Also

\bea
S^{(B)}_{0,1,2}&=&\frac{1}{k^2}J_-\bar{\phi}J_- \phi \nn\\
&+& \frac{1}{k^2}\cf_{3-}\bar{\phi}J_- \phi +\frac{1}{k^2}\cf_{3-}\bar{\phi}\cf_{3-}\phi+ \frac{1}{k^3}f\Lambda_-J_-\bar{\phi}J_-\phi \nn\\
&+&\frac{1}{k^3} \cf_{3-}\bar{\phi}f\Lambda_-J_- \phi + \frac{1}{k^3}f\Lambda_-\cf_{3-}\bar{\phi}J_-\phi+\frac{1}{k^4}f^2(\Lambda_-)^2J_-\bar{\phi}J_- \phi \nn\\
\eea
The above expression contains the zeroth, first and second order corresponding to each line. Terms begin to proliferate starting at third order where we have contribution from products of first and second order terms in addition to intrinsically third order terms in $A_3$:

\bea
S^{(B)}_3&=&\frac{1}{k^4}f^2 (\L_-)^2\cf_{3-}\bar{\phi}J_- \phi + \frac{1}{k^5}f^3(\L_-)^3J_-\bar{\phi}J_- \phi  \nn\\
&+&\frac{1}{k^3} \cf_{3-}\bar{\phi}f\Lambda_-\cf_{3-} \phi+\frac{1}{k^4}\cf_{3-}\bar{\phi}f^2(\Lambda_-)^2J_- \phi \nn\\
&+&\frac{1}{k^4} f\, \L_- J_-\bar{\phi}f\Lambda_-\cf_{3-} \phi + \frac{1}{k^5}f\Lambda_-J_-\bar{\phi}f^2(\Lambda_-)^2J_- \phi \nn\\
\eea
Let us now present the explicit form of some of the schematic terms above with all the indices and momentum factors:

\bea
 \frac{1}{k^3}f\Lambda_-J_-\bar{\phi}J_-\phi &\Longrightarrow& \frac{1}{4}\int\frac{d^3p d^3q d^3r d^3p_1 d^3\omega_1 d^3\omega_2 }{(2\pi)^{18}}\frac{2\pi i}{k p_-}\frac{2\pi i}{k q_-}\frac{2\pi i}{k (p-p_1)_-}(2\omega_1+p-p_1)_-(2\omega_2+q)_-\nonumber\\
&&\lbrack \bar{\phi}(r)\phi(\omega_1+p-p_1)\rbrack
\lbrack\bar{\phi}(-\omega_1)\L(p_1)\phi(\omega_2+q)\rbrack\lbrack\bar{\phi}(-\omega_2)\phi(-p-q-r)\rbrack
\nonumber\\
&&-\frac{1}{4}\int\frac{d^3p d^3q d^3r d^3p_1 d^3\omega_1 d^3\omega_2 }{(2\pi)^{18}}
\frac{2\pi i}{k p_-}\frac{2\pi i}{k q_-}\frac{2\pi i}{k (p-p_1)_-}(2\omega_1+p-p_1)_-(2\omega_2+q)_-
\nonumber\\
&&\lbrack \bar{\phi}(-\omega_1)\phi(\omega_2+q)\rbrack
\lbrack\bar{\phi}(r)\L(p_1)\phi(\omega_1+p-p_1)\rbrack
\lbrack\bar{\phi}(-\omega_2)\phi(-p-q-r)\rbrack
\nonumber\\
%\eea
%\bea
& & \nonumber\\
 \frac{1}{k^2}\cf_{3-}\bar{\phi}J_- \phi &\Longrightarrow &- \frac{1}{2}\int\frac{d^3p d^3q d^3r d^3\omega_1  }{(2\pi)^{12}}\frac{2\pi i}{k p_-}\frac{2\pi i}{k q_-}
(2\omega_1+p)_-\nonumber\\
&&
\lbrack\bar{\phi}(-\omega_1)\cf_{3-}(q)\phi(-p-q-r)\rbrack\lbrack\bar{\phi}(r)\phi(\omega_1+p)\rbrack \nonumber \\
&&
 - \frac{1}{2}\int\frac{d^3p d^3q d^3r d^3\omega_1  }{(2\pi)^{12}}\frac{2\pi i}{k p_-}\frac{2\pi i}{k q_-}
(2\omega_1+q)_- \nonumber \\
&&\lbrack\bar{\phi}(r)
\cf_{3-}(p) \phi(\omega_1+q)\rbrack
\lbrack\bar{\phi}(-\omega_1)\phi(-p-q-r)\rbrack
\eea
Also:
\bea
&&\frac{1}{k^3} \cf_{3-}\bar{\phi}f\Lambda_-J_- \phi + \frac{1}{k^3}f\Lambda_-\cf_{3-}\bar{\phi}J_-\phi  \Longrightarrow \nonumber \\ &&\nonumber\\
&& \frac{1}{2}\int\frac{d^3p d^3q d^3r d^3p_1 d^3\omega_1  }{(2\pi)^{15}}\frac{2\pi i}{k p_-}\frac{2\pi i}{k q_-}\frac{2\pi i}{k (p-p_1)_-}(2\omega_1+p-p_1)\nonumber \\
&&\lbrack \bar{\phi}(r)\phi(\omega_1+p-p_1)\rbrack
\lbrack\bar{\phi}(-\omega_1)\L(p_1)
\cf_{3-}(q)\phi(-p-q-r)\rbrack\nonumber\\
&&\nonumber\\
&&+\frac{1}{2}\int\frac{d^3p d^3q d^3r d^3p_1 d^3\omega_1  }{(2\pi)^{15}}\frac{2\pi i}{k p_-}\frac{2\pi i}{k q_-}\frac{2\pi i}{k (p-p_1)_-}(2\omega_1+q)
 \nonumber\\
&&\lbrack \bar{\phi}(-\omega_1) \phi(-p-q-r)\rbrack \lbrack\bar{\phi}(r)\cf_{3-}(p-p_1) \L(p_1)\phi(\omega_1+q\rbrack \nonumber \\
&&\nonumber\\
&&-\frac{1}{2}\int\frac{d^3p d^3q d^3r d^3p_1 d^3\omega_1  }{(2\pi)^{15}}\frac{2\pi i}{k p_-}\frac{2\pi i}{k q_-}\frac{2\pi i}{k (p-p_1)_-}(2\omega_1+p-p_1)\nonumber\\
&& \lbrack \bar{\phi}(r)\L(p_1)\phi(\omega_1+p-p_1)\rbrack
\lbrack\bar{\phi}(-\omega_1)\cf_{3-}(q)\phi(-p-q-r)\rbrack\nonumber\\
&&\\
&&  -\frac{1}{2}\int\frac{d^3p d^3q d^3r d^3p_1 d^3\omega_1  }{(2\pi)^{15}}
\frac{2\pi i}{k p_-}\frac{2\pi i}{k q_-}\frac{2\pi i}{k (p-p_1)_-}
(2\omega_1+q)\nonumber\\
&& \lbrack \bar{\phi}(-\omega_1)
\phi(-p-q-r)\rbrack
\lbrack\bar{\phi}(r)
\L(p_1)\cf_{3-}(p-p_1)\phi(\omega_1+q)\rbrack \nonumber\\
&&\frac{1}{4}\int\frac{d^3p d^3q d^3r d^3q_1 d^3\omega_1 d^3\omega_2 }{(2\pi)^{18}}\frac{2\pi i}{k p_-}\frac{2\pi i}{k q_-}\frac{2\pi i}{k (q-q_1)_-}(2\omega_1+p)
(2\omega_2+q-q_1) \nonumber \\
&& \lbrack \bar{\phi}(r)
\phi(\omega_1+p)\rbrack
\lbrack\bar{\phi}(-\omega_2)
\L(q_1)\phi(-p-q-r)\rbrack\lbrack\bar{\phi}(-\omega_1)\phi(\omega_2+q-q_1)\rbrack\nonumber\\
&&\nonumber\\
&&+\frac{1}{2}\int\frac{d^3p d^3q d^3r d^3q_1 d^3\omega_1  }{(2\pi)^{15}}\frac{2\pi i}{k p_-}\frac{2\pi i}{k q_-}\frac{2\pi i}{k (q-q_1)_-}
\nonumber\\
&&(2\omega_1+p)
\lbrack \bar{\phi}(r)
\phi(\omega_1+p)\rbrack
\lbrack\bar{\phi}(-\omega_1)\cf_{3-}(q-q_1)\L(q_1)
\phi(-p-q-r)\rbrack + \ldots \nonumber\\
&&\nonumber
\eea

When we plug in the  perturbative solution for $A^a_3(p)$ in the action for the fundamental bosons and fermions, we observe that at nth order the types of  terms involving Lagrange multiplier field
$\L$ are exhausted by the following combinations in momentum space (Note that we will use Einstein summation convention):

For the fundamental boson action:
\bea
&\lbrack\bar{\phi}T^{a_1}T^{a_2}T^{a_3}...T^{a_n}\phi \L^{a_1}\L^{a_2}\L^{a_3}...\L^{a_n}\rbrack
\eea
\bea
&\lbrack\bar{\phi}T^{a_1}T^{a_2}T^{a_3}...T^{a_n}\phi \L^{a_1}\L^{a_2}\L^{a_3}...\L^{a_n}\rbrack
(\bar{\phi}\phi)
\nonumber\\
&\lbrack\bar{\phi}T^{a_1}T^{a_2}T^{a_3}...T^{a_{n-m}}\phi \L^{a_1}\L^{a_2}\L^{a_3}...\L^{a_{n-m}}\rbrack
\lbrack\bar{\phi}T^{a_{n-m+1}}...T^{a_{n}}\phi \L^{a_{n-m+1}}...\L^{a_{n}}\rbrack
\nonumber\\
\eea

\bea
&\lbrack\bar{\phi}T^{a_1}T^{a_2}T^{a_3}...T^{a_n}\phi\L^{a_1}\L^{a_2}\L^{a_3}...\L^{a_n}\rbrack
(\bar{\phi}\phi)
(\bar{\phi}\phi)
\nonumber\\
&\lbrack\bar{\phi}T^{a_1}T^{a_2}T^{a_3}...T^{a_{n-m}}\phi \L^{a_1}\L^{a_2}\L^{a_3}...\L^{a_{n-m}}\rbrack
\lbrack\bar{\phi}T^{a_{n-m+1}}...T^{a_{n}}\phi \L^{a_{n-m+1}}...\L^{a_{n}}\rbrack(\bar{\phi}\phi)
\nonumber\\
&\lbrack\bar{\phi}T^{a_1}T^{a_2}T^{a_3}...T^{a_{n-l}}\phi \L^{a_1}\L^{a_2}\L^{a_3}...\L^{a_{n-l}}\rbrack
\lbrack\bar{\phi}T^{a_{n-l+1}}...T^{a_{n}}\phi \L^{a_{n-l+1}}...\L^{a_{m}}\rbrack\times
\nonumber\\
&\lbrack\bar{\phi}T^{a_{m+1}}...T^{a_{n}}\phi\L^{a_{m+1}}...\L^{a_{n}}\rbrack
\eea
For the fundamental fermion action:
\bea
&\lbrack\bar{\psi}\gamma_3T^{a_1}T^{a_2}T^{a_3}...T^{a_n}\psi \L^{a_1}\L^{a_2}\L^{a_3}...\L^{a_n}\rbrack
\eea
\bea
&\lbrack\bar{\psi}\gamma_{-}T^{a_1}T^{a_2}T^{a_3}...T^{a_n}\phi\L^{a_1}\L^{a_2}\L^{a_3}...\L^{a_n}\rbrack
(\bar{\psi}\psi)
\nonumber\\
&\lbrack\bar{\psi}\gamma_{-}T^{a_1}T^{a_2}T^{a_3}...T^{a_{n-m}}\psi\L^{a_1}\L^{a_2}\L^{a_3}...\L^{a_{n-m}}\rbrack
\lbrack\bar{\psi}T^{a_{n-m+1}}...T^{a_{n}}\psi \L^{a_{n-m+1}}...\L^{a_{n}}\rbrack
\nonumber\\
&\lbrack\bar{\psi}\gamma_{-}T^{a_1}T^{a_2}T^{a_3}...T^{a_{n-m}}\psi\L^{a_1}\L^{a_2}\L^{a_3}...\L^{a_{n-m}}\rbrack
\lbrack\bar{\psi}\gamma_3T^{a_{n-m+1}}...T^{a_{n}}\psi \L^{a_{n-m+1}}...\L^{a_{n}}\rbrack
\nonumber\\
\eea
where $T^a$ is the generator of $SU(N)$ in the fundamental representation.\\
The universal piece:
\bea
\frac{1}{k}\cf_{3-}\cf_{+-}
\eea

\subsection{Large N Analysis:}

First we prove the following  identity
\bea
(T^{a_1}T^{a_2}T^{a_3}...T^{a_n})={\mathcal{O}}\left(\frac{1}{N}\right)+F^{a_1a_2b_1}F^{b_1a_3b_2}F^{b_2a_4b_3}...F^{b_{n-1} a_{n} b_n}T^{b_n}
\eea
where $a_1,....,a_n,b_1,...,b_n$ run from $1$ to $N^2-1$ ,
$F^{abc}\equiv \frac{1}{2}(d^{abc}+ i f^{abc})$ with $d^{abc},f^{abc}$ being structure constants of
$SU(N)$ Lie algebra.\\
Proof :
First few cases are  manifestly true
\bea
&T^{a_1}T^{a_2}=\frac{\delta^{a_1a_2}}{N}+F^{a_1a_2b_1}T^{b_1}
\nonumber\\
&T^{a_1}T^{a_2}T^{a_3}=\frac{\delta^{a_1a_2}T^{a_3}}{N}+\frac{F^{a_1a_2a_3}}{N}+F^{a_1a_2b_1}F^{b_1a_3b_2}T^{b_2}
\nonumber\\
&T^{a_1}T^{a_2}T^{a_3}T^{a4}=\frac{\delta^{a_1a_2}\delta^{a_3a_4}}{N^2}+\frac{\delta^{a_1a_2}F^{a_3a_4b_3}T^{b_3}}{N}+\frac{F^{a_1a_2a_3}T^{a_4}}{N}+\frac{F^{a_1a_2b_1}F^{b_1a_3a_4}}{N}
\nonumber\\
&+ F^{a_1a_2b_1}F^{b_1a_3b_2}F^{b_2a_4b_3}T^{b_3}
\nonumber\\
\eea
{\bf Odd n}:\\
Assume that  the assertion is true for $T^{a_1}T^{a_2}T^{a_3}...T^{a_{n-1}}$ i.e.
\bea
T^{a_1}T^{a_2}T^{a_3}...T^{a_{n-1}}={\mathcal{O}}\left(\frac{1}{N^{\frac{n-1}{2}}}\right)+{\mathcal{O}}\left(\frac{1}{N^{\frac{n-1}{2}-1}}\right)+...+{\mathcal{O}}\left(\frac{1}{N}\right)
\nonumber\\
+F^{a_1a_2b_1}F^{b_1a_3b_2}F^{b_2a_4b_3}...F^{b_{n-2} a_{n-1} b_{n-1}}T^{b_{n-1}}
\eea
then
\bea
T^{a_1}T^{a_2}T^{a_3}...T^{a_{n}} &=&\lbrack(T^{a_1}T^{a_2}T^{a_3}...T^{a_{n-1}}T^{a_{n}})\rbrack
\nonumber\\ &
=&(\lbrack {\mathcal{O}}\left(\frac{1}{N^{\frac{n-1}{2}}}\right)+{\mathcal{O}}\left(\frac{1}{N^{\frac{n-1}{2}-1}}\right)+...+{\mathcal{O}}\left(\frac{1}{N}\right)
\nonumber\\
&+&F^{a_1a_2b_1}F^{b_1a_3b_2}F^{b_2a_4b_3}...F^{b_{n-2} a_{n-1} b_{n-1}}T^{b_{n-1}}\rbrack T^{a_n})
\nonumber\\&
=& {\mathcal{O}}\left(\frac{1}{N^{\frac{n-1}{2}}}\right)+{\mathcal{O}}\left(\frac{1}{N^{\frac{n-1}{2}-1}}\right)+...+{\mathcal{O}}\left(\frac{1}{N}\right)
\nonumber\\
&+&F^{a_1a_2b_1}F^{b_1a_3b_2}F^{b_2a_4b_3}...F^{b_{n-1} a_{n} b_n}T^{b_n}
\eea
where in going from second to third line we used the special case $n=2$ of the assertion .\\
{\bf Even n}:\\
Assume that  the assertion is true for $T^{a_1}T^{a_2}T^{a_3}...T^{a_{n-1}}$ i.e.
\bea
T^{a_1}T^{a_2}T^{a_3}...T^{a_{n-1}}={\mathcal{O}}\left(\frac{1}{N^{\frac{n-2}{2}}}\right)T^{a_{n-1}}+{\mathcal{O}}\left(\frac{1}{N^{\frac{n-1}{2}-1}}\right)+...+{\mathcal{O}}\left(\frac{1}{N}\right)
\nonumber\\
+F^{a_1a_2b_1}F^{b_1a_3b_2}F^{b_2a_4b_3}...F^{b_{n-2} a_{n-1} b_{n-1}}T^{b_{n-1}}
\eea
then
\bea
T^{a_1}T^{a_2}T^{a_3}...T^{a_{n}} &=&\lbrack(T^{a_1}T^{a_2}T^{a_3}...T^{a_{n-1}}T^{a_{n}})\rbrack
\nonumber\\ &
=&\left(\left[ {\mathcal{O}}\left(\frac{1}{N^{\frac{n-2}{2}}}\right)T^{a_{n-1}}+{\mathcal{O}}\left(\frac{1}{N^{\frac{n-1}{2}-1}}\right)+...\right.\right.\nonumber\\
&+&\left. \left. {\mathcal{O}}\left(\frac{1}{N}\right)
%\nonumber\\
%&+&\left.
+ F^{a_1a_2b_1}F^{b_1a_3b_2}F^{b_2a_4b_3}...F^{b_{n-2} a_{n-1} b_{n-1}}T^{b_{n-1}}\right] T^{a_n}\right)
\nonumber\\&
=& {\mathcal{O}}\left(\frac{1}{N^{\frac{n}{2}}}\right)+{\mathcal{O}}\left(\frac{1}{N^{\frac{n}{2}-1}}\right)+...+{\mathcal{O}}\left(\frac{1}{N}\right)
\nonumber\\
&+&F^{a_1a_2b_1}F^{b_1a_3b_2}F^{b_2a_4b_3}...F^{b_{n-1} a_{n} b_n}T^{b_n}
\eea
where in going from second to third line we again used the special case $n=2$ of the assertion
Q.E.D.\\
This shows that in the large $N$ limit
\bea
\lim_{N{\to\infty}}(T^{a_1}T^{a_2}T^{a_3}...T^{a_{n}})=F^{a_1a_2b_1}F^{b_1a_3b_2}F^{b_2a_4b_3}...F^{b_{n-1} a_{n} b_n}T^{b_n}
\eea
Now using this assertion we can see that
\bea
\lbrack\bar{\phi}T^{a_1}T^{a_2}T^{a_3}...T^{a_n}\phi \L^{a_1}\L^{a_2}\L^{a_3}...\L^{a_n}\rbrack &=&\bar{\phi}T^{b_{n}}\phi
 \lbrack(F^{a_1a_2b_1}F^{b_1a_3b_2}F^{b_2a_4b_3}...F^{b_{n-2} a_{n-1} b_{n-1}}
 \nonumber\\&
& F^{b_{n-1} a_{n} b_{n}}
\L^{a_1}\L^{a_2}\L^{a_3}...\L^{a_n})\rbrack
\eea
but $\bar{\phi}T^{b_{n}}\phi$ does not belong to the singlet sector, hence this term is subleasing in the large N-limit.\\
Next the expression
\bea
\lbrack\bar{\phi}T^{a_1}T^{a_2}T^{a_3}...T^{a_{n-m}}\phi \L^{a_1}\L^{a_2}\L^{a_3}...\L^{a_{n-m}}\rbrack
\lbrack\bar{\phi}T^{a_{n-m+1}}...T^{a_{n}}\phi \L^{a_{n-m+1}}...\L^{a_{n}}\rbrack
\eea
 reduces in the limit $N\to \infty$ to
 \bea
 F^{a_1a_2b_1}...F^{b_{n-m-1}a_{n-m}b_{n-m}} F^{a_{n-m+1}a_{n-m+2}b_{n-m+1}}...F^{b_{n-1}a_nb_n}\bar{\phi}T^{b_{n-m}}\phi\bar{\phi}T^{b_n}\phi
 \eea
 Using eq. $(C.10)$ it is easy to see that
 \bea
&F^{a_1a_2b_1}...F^{b_{n-m-1}a_{n-m}b_{n-m}} F^{a_{n-m+1}a_{n-m+2}b_{n-m+1}}...F^{b_{n-1}a_nb_n}\bar{\phi}T^{b_{n-m}}\phi\bar{\phi}T^{b_n}\phi
\nonumber\\
&\sim {\mathcal{O}}\left(\frac{1}{N}\right)+{\mathcal{O}}\left(\frac{1}{N^2}\right)
 \eea
This shows that  expressions of the form
\bea
\bar{\phi}T^{a_1}...T^{a_n-l}\phi\bar{\phi}T^{a_{n-l+1}}...T^{a_{n-m}}\phi\bar{\phi}T^{a_{n-m}}...T^{a_n}\phi
\eea
will be even more subleading in the large N limit.\\
The analysis of the fermionic terms similarly goes through.
\\
Let us analyze $S[\L]$
\bea
S[\L]=-\int d^3x N^{ab} \cf_{-+}^a\cf_{-3}^b -\tr\log\left(\frac{k}{4\pi}\delta^{ab}\partial_- - f^{ab}{}_c \Lambda_-^c\right)
\eea
now suppressing the indices
\bea
\tr\log\left(\frac{k}{4\pi}\delta^{ab}\partial_- - f^{ab}{}_c \Lambda_-^c\right)&=& \tr \log\left(\frac{k}{4\pi}I \partial_- - f \Lambda_-\right)\nonumber\\
&=&\tr\log\left(\frac{4\pi}{k \partial_-}\right)+\tr\log\left(I-\frac{4\pi  f \L_-}{k \partial_-}\right)
\eea
and  making a series expansion
\bea
\tr\log\left(I-\frac{4\pi  f \L_-}{k \partial_-}\right)&=&-\tr\sum^{\infty}_{n=1}\frac{(-1)^n(-\frac{ 4\pi  f\L_-}{k\partial_-})^n}{n }\nonumber\\
\eea
We have analyzed terms of the form $f^n\Lambda^n/k^n$. For odd $n$,  in the large $N$ limit, their contribution vanishes. For even $n$ one can explicitly demonstrate that the answer is only a function of $\lambda=N/k$. Therefore the $N\leftrightarrow k$ symmetry is preserved in this expansion. 
\section{Diagramatic expansion of the effective actions upto $\mathcal{O}(\frac{1}{N})$}\label{Sec:diagrammatics}
\subsection{One Point amplitude }
%Note that $\cf_{-3}(p)=i p_-\L_3(p)-i p_3\L_-(p),\cf_{+-}(p)=i p_+\L_-(p)-i p_-\L_+(p)$ .
It is important to keep in mind that  there is an overall factor of N for each vertex, because both and bosons and fermions are in the fundamental representations of $U(N)$. This is accounted for as an overall factor of $N$ in the effective actions. 
Also the bosonic and fermonic propagators are large $N$ exact.
\subsection*{Bosonic theory}
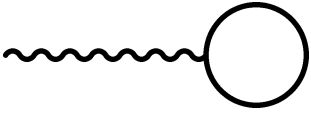
\begin{fmffile}{bq6}
\begin{fmfgraph}(40,30)
\fmfpen{thick} 
\fmfleft{i}
\fmfright{o} 
\fmf{photon}{i,v1}
%\fmf{photon}{v2,o}
\fmf{plain,left}{v1,o,v1}
% \fmf{plain,left=95}{v,v}
  %\fmfdot{v}
  \end{fmfgraph}
\end{fmffile}

\bea
&=&\tr{(T^a)}\frac{2 \pi}{k}\int  \frac{d^3p}{(2\pi)^3} \frac{d^3p_1}{(2\pi)^3}\frac{1}{p_1^2 p_-}(\cf_{-3}^a(p)(2 (p_1)_3+p_3)+\frac{1}{2}\cf_{+-}^a(p)(2 (p_1)_-+p_-))\nonumber\\
\eea

\subsection*{Fermionic theory}
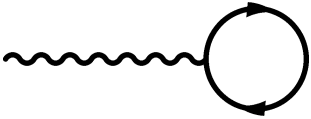
\begin{fmffile}{qp4}
\begin{fmfgraph}(40,30)
\fmfpen{thick} 
\fmfleft{i}
\fmfright{o} 
\fmf{photon}{i,v1}
%\fmf{photon}{v2,o}
\fmf{fermion,left}{v1,o,v1}
% \fmf{plain,left=95}{v,v}
  %\fmfdot{v}
  \end{fmfgraph}
\end{fmffile}
\bea
&=&-\tr{(T^a)}\frac{2 \pi}{k}\int  \frac{d^3p}{(2\pi)^3} \frac{d^3p_1}{(2\pi)^3}\frac{1}{p_1^2 p_-}(2\cf_{-3}^a(p) (p_1)_3+2\cf_{+-}^a(p) (p_1)_-)\nonumber\\
\eea
These amplitudes for Feynman diagrams corresponding to one-point functions vanish due to symmetry properties of the integrands and the regularization scheme \cite{Giombi:2011kc} we are using in which
\bea
\int  \frac{d^3p}{(2\pi)^3} \frac{1}{p^2}=0
\eea

\subsection*{Scaling with number of colors N}
If the Lagrange multiplier field $\L^a$ and the Chern Simons gauge field $A^a$ have the  same large $N$ scaling.\\
Under this scaling, the  $1$-point  amplitudes for $\L^a$ are of $\mathcal{O}(\frac{1}{N})$.

\subsection{Two Point amplitude}
\subsection*{ Bosonic theory}

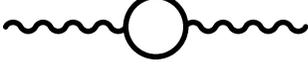
\begin{fmffile}{qb9}
\begin{fmfgraph}(40,30)
\fmfpen{thick} 
\fmfleft{i}
\fmfright{o} 
\fmf{photon}{i,v1}
\fmf{photon}{v2,o}
\fmf{plain,left}{v1,v2,v1}
% \fmf{plain,left=95}{v,v}
  %\fmfdot{v}
  \end{fmfgraph}
\end{fmffile}
\bea
&=&
\frac{1}{2}\frac{(2 \pi)^2}{k^2}\int  \frac{d^3p}{(2\pi)^3} \frac{d^3p_1}{(2\pi)^3}\frac{1}{p_-^2}\frac{1}{(p_1+p)^2 p_1^2}\nonumber\\
&\times& \tr \bigg[\bigg(\cf_{+-}^b(p) T^b\frac{1}{2}(2p_1+p)_-+\cf_{-3}^b(p)T^b(2p_1+p)_3\bigg)\nonumber\\
&\times&\bigg(\cf_{+-}^a(-p) T^a (p_1)_-+\cf_{-3}^a(-p)T^a (2p_1)_3\bigg)\bigg]\nonumber\\
\eea

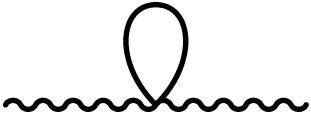
\begin{fmffile}{qb8}
\begin{fmfgraph}(40,30)
\fmfpen{thick} 
\fmfleft{i}
\fmfright{o} 
\fmf{photon}{i,v}
\fmf{plain}{v,v}
\fmf{photon}{v,o}
% \fmf{plain,left=95}{v,v}
  %\fmfdot{v}
  \end{fmfgraph}
\end{fmffile}

\bea
&=&
\frac{(2 \pi)^2}{k^2}\int  \frac{d^3p}{(2\pi)^3} \frac{d^3p_1}{(2\pi)^3}\frac{1}{p_-^2}\frac{1}{p_1^2}\nonumber\\
&\times& \tr \bigg[\bigg(\cf_{-3}^a(p)\cf_{-3}^b(-p)T^aT^b\bigg)\bigg]\nonumber\\
\eea
This tadpole diagram evaluates to zero.
\subsection*{ Fermionic theory}
%%%%%%
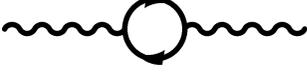
\begin{fmffile}{q9}
\begin{fmfgraph}(40,30)
\fmfpen{thick} 
\fmfleft{i}
\fmfright{o} 
\fmf{photon}{i,v1}
\fmf{photon}{v2,o}
\fmf{fermion,left}{v1,v2,v1}
% \fmf{plain,left=95}{v,v}
  %\fmfdot{v}
  \end{fmfgraph}
\end{fmffile}
%%%%%%
\bea
&=&-\frac{1}{2}\frac{(2 \pi)^2}{k^2}\int  \frac{d^3p}{(2\pi)^3} \frac{d^3p_1}{(2\pi)^3}\frac{1}{p_-^2}\frac{1}{(p_1+p)^2 (p_1)^2} \tr \bigg[\bigg(\cf_{+-}^b(p) T^b\gamma_-+\cf_{-3}^b(p)T^b\gamma_3\bigg)\nonumber\\
&\times&\bigg(-i (\gamma_- (p_1+p)_+(1-\lambda^2)+(p_1+p)_-\gamma_++(p_1+p)_3\gamma_3)+\lambda(2 (p_1+p)_+(p_1+p)_-)^{\frac{1}{2}}\bigg)\nonumber\\
&\times&\bigg(\cf_{+-}^a(-p) T^a \gamma_-+\cf_{-3}^a(-p)T^a \gamma_3\bigg)\nonumber\\
&\times&\bigg(-i (\gamma_- (p_1)_+(1-\lambda^2)+(p_1)_-\gamma_++(p_1)_3\gamma_3)+\lambda(2 (p_1)_+(p_1)_-)^{\frac{1}{2}}\bigg)
\bigg]\nonumber\\
\eea
\subsection*{Scaling with number of colors N}
$2$-point  amplitudes of $\L^a$  are of $\mathcal{O}(1)$.
\subsection{Three Point amplitude}
\subsection*{Bosonic Theory}
\begin{fmffile}{wb10}
\begin{fmfgraph}(40,30)
\fmfpen{thick} 
\fmfleft{i1,i2} 
\fmfright{o1}
\fmf{plain}{v1,v3}
\fmf{plain}{v3,v2}
\fmf{plain}{v2,v1}
  \fmf{photon}{i1,v1}
   \fmf{photon}{v3,o1}
   \fmf{photon}{i2,v2}
\fmfdotn{v}{3}
  %\fmfdot{v}
  \end{fmfgraph}
\end{fmffile}

\bea
&=&\frac{1}{3}\frac{(2 \pi)^3}{k^3}\int  \frac{d^3p}{(2\pi)^3} \frac{d^3q}{(2\pi)^3} \frac{d^3p_1}{(2\pi)^3}\frac{1}{p_-q_-(-p-q)_-}\frac{1}{(p_1+p)^2 (p_1+p+q)^2(p_1)^2}\nonumber\\
&\times& \tr \bigg[\bigg(\cf_{+-}^b(p) T^b\frac{1}{2}(2p_1+p)_-+\cf_{-3}^b(p)T^b(2p_1+p)_3\bigg)\nonumber\\
&\times&\bigg(\cf_{+-}^c(q) T^c\frac{1}{2}(2p_1+p+q)_-+\cf_{-3}^c(q)T^c(2p_1+p+q)_3\bigg)\nonumber\\
&\times&\bigg(\cf_{+-}^a(-p-q) T^a (p_1)_-+\cf_{-3}^a(-p-q)T^a (2p_1)_3\bigg)\bigg]\nonumber\\
\eea

\begin{fmffile}{wb12}
\begin{fmfgraph}(40,30)
\fmfpen{thick} 
\fmfleft{i1,i2} 
\fmfright{o1}
\fmf{plain,left=0.3,tension=0.5}{v1,v2}
\fmf{plain,left=0.3,tension=0.5}{v2,v1}
  \fmf{photon}{i1,v1}
   \fmf{photon}{v2,o1}
   \fmf{photon}{i2,v1}
\fmfdotn{v}{2}
  %\fmfdot{v}
  \end{fmfgraph}
\end{fmffile}

\bea
&=&\frac{1}{2}\frac{(2 \pi)^3}{k^3}\int  \frac{d^3p}{(2\pi)^3} \frac{d^3q}{(2\pi)^3} \frac{d^3p_1}{(2\pi)^3}\frac{1}{p_-q_-(-p-q)_-}\frac{1}{ (p_1+p+q)^2(p_1)^2}\nonumber\\
&\times& \tr \bigg[\bigg(\cf_{-3}^a(p)\cf_{-3}^b(q)T^aT^b\bigg)\nonumber\\
&\times&\bigg(\cf_{+-}^c(-p-q) T^c\frac{1}{2}(2p_1+p+q)_-+\cf_{-3}^c(-p-q)T^c(2p_1+p+q)_3\bigg)\bigg]\nonumber\\
\eea

\subsection*{Fermionic Theory}
\begin{fmffile}{w10}
\begin{fmfgraph}(40,30)
\fmfpen{thick} 
\fmfleft{i1,i2} 
\fmfright{o1}
\fmf{fermion}{v1,v3}
\fmf{fermion}{v3,v2}
\fmf{fermion}{v2,v1}
  \fmf{photon}{i1,v1}
   \fmf{photon}{v3,o1}
   \fmf{photon}{i2,v2}
\fmfdotn{v}{3}
  %\fmfdot{v}
  \end{fmfgraph}
\end{fmffile}
\bea
&=&-\frac{1}{3}\frac{(2 \pi)^3}{k^3}\int  \frac{d^3p}{(2\pi)^3} \frac{d^3q}{(2\pi)^3} \frac{d^3p_1}{(2\pi)^3}\frac{1}{p_-q_-(-p-q)_-}\frac{1}{(p_1+p)^2 (p_1+p+q)^2(p_1)^2}\nonumber\\
&\times& \tr \bigg[\bigg(\cf_{+-}^a(p) T^a\gamma_-+\cf_{-3}^a(p)T^a\gamma_3\bigg)\nonumber\\
&\times&\bigg(-i (\gamma_- (p_1+p)_+(1-\lambda^2)+(p_1+p)_-\gamma_++(p_1+p)_3\gamma_3)+\lambda(2 (p_1+p)_+(p_1+p)_-)^{\frac{1}{2}}\bigg)\nonumber\\
&\times&\bigg(\cf_{+-}^b(q) T^b\gamma_-+\cf_{-3}^b(q)T^b\gamma_3\bigg)\nonumber\\
&\times&\bigg(-i (\gamma_- (p_1+p+q)_+(1-\lambda^2)+(p_1+p+q)_-\gamma_++(p_1+p+q)_3\gamma_3)\nonumber\\
&+&\lambda(2 (p_1+p+q)_+(p_1+p+q)_-)^{\frac{1}{2}}\bigg)\nonumber\\
&\times&\bigg(\cf_{+-}^c(-p-q) T^c \gamma_-+\cf_{-3}^c(-p-q)T^c \gamma_3\bigg)\nonumber\\
&\times&\bigg(-i (\gamma_- (p_1)_+(1-\lambda^2)+(p_1)_-\gamma_++(p_1)_3\gamma_3)+\lambda(2 (p_1)_+(p_1)_-)^{\frac{1}{2}}\bigg)
\bigg]\nonumber\\
\eea
\subsection*{Scaling with number of colors N}
$3$-point  amplitudes are of $\mathcal{O}(\frac{1}{N})$. Note that  $\tr(T^aT^bT^c)=i f^{abc}+d^{abc}$ and $\L^a\L^b\L^c f^{abc}$ will of order less than $\frac{1}{N}$. Therefore in three point functions only $\L^a\L^b\L^c d^{abc}$  will enter at $\mathcal{O}(\frac{1}{N})$.
\subsection{Four Point amplitude}
\subsection*{Bosonic Theory}
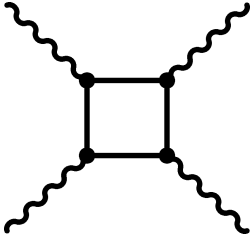
\begin{fmffile}{wboson6}
\begin{fmfgraph}(40,30)
\fmfpen{thick} 
\fmfleft{i1,i2} 
\fmfright{o1,o2}
\fmf{plain}{v1,v3}
\fmf{plain}{v3,v4}
\fmf{plain}{v4,v2}
\fmf{plain}{v2,v1}
  \fmf{photon}{i1,v1}
   \fmf{photon}{v3,o1}
   \fmf{photon}{i2,v2}
   \fmf{photon}{v4,o2}
\fmfdotn{v}{4}
  %\fmfdot{v}
  \end{fmfgraph}
\end{fmffile}
\bea
&=&\frac{1}{4}\frac{(2 \pi)^4}{k^4}\int  \frac{d^3p}{(2\pi)^3} \frac{d^3q}{(2\pi)^3} \frac{d^3s}{(2\pi)^3} \frac{d^3p_1}{(2\pi)^3}\frac{1}{p_-q_-s_-(-p-q-s)_-}\frac{1}{(p_1+p)^2 (p_1+p+q)^2}\nonumber\\
&\times&\frac{1}{(p_1+p+q+s)^2(p_1)^2}\nonumber\\
&\times& \tr \bigg[\bigg(\cf_{+-}^b(p) T^b\frac{1}{2}(2p_1+p)_-+\cf_{-3}^b(p)T^b(2p_1+p)_3\bigg)\nonumber\\
&\times&\bigg(\cf_{+-}^b(q) T^b\frac{1}{2}(2p_1+p+q)_-+\cf_{-3}^b(q)T^b(2p_1+p+q)_3\bigg)\nonumber\\
&\times&\bigg(\cf_{+-}^b(s) T^b\frac{1}{2}(2p_1+p+q+s)_-+\cf_{-3}^b(q)T^b(2p_1+p+q+s)_3\bigg)\nonumber\\
&\times&\bigg(\cf_{+-}^a(-p-q-s) T^a (p_1)_-+\cf_{-3}^a(-p-q-s)T^a (2p_1)_3\bigg)\bigg]\nonumber\\
\eea

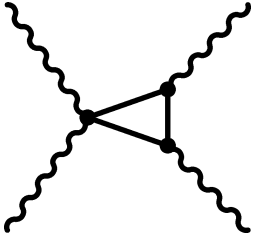
\vspace{0.6mm}

\begin{fmffile}{wboson8}
\begin{fmfgraph}(40,30)
\fmfpen{thick} 
\fmfleft{i1,i2} 
\fmfright{o1,o2}
\fmf{plain}{v1,v3}
\fmf{plain}{v3,v2}
\fmf{plain}{v2,v1}
  \fmf{photon}{i1,v1}
   \fmf{photon}{v3,o1}
   \fmf{photon}{i2,v1}
   \fmf{photon}{v2,o2}
\fmfdotn{v}{3}
  %\fmfdot{v}
  \end{fmfgraph}
\end{fmffile}
\bea
&=&\frac{1}{3}\frac{(2 \pi)^4}{k^4}\int  \frac{d^3p}{(2\pi)^3} \frac{d^3q}{(2\pi)^3} \frac{d^3s}{(2\pi)^3} \frac{d^3p_1}{(2\pi)^3}\frac{1}{p_-q_-s_-(-p-q-s)_-}\frac{1}{(p_1+p+q)^2}\nonumber\\
&\times&\frac{1}{(p_1+p+q+s)^2(p_1)^2}\nonumber\\
&\times& \tr \bigg[\bigg(\cf_{-3}^a(p)\cf_{-3}^b(q)T^aT^b\bigg)\nonumber\\
%&\times&\bigg(\cf_{+-}^b(q) T^b\frac{1}{2}(2p_1+p+q)_-+\cf_{-3}^b(q)T^b(2p_1+p+q)_3\bigg)\nonumber\\
&\times&\bigg(\cf_{+-}^c(s) T^c\frac{1}{2}(2p_1+p+q+s)_-+\cf_{-3}^c(q)T^c(2p_1+p+q+s)_3\bigg)\nonumber\\
&\times&\bigg(\cf_{+-}^d(-p-q-s) T^d (p_1)_-+\cf_{-3}^d(-p-q-s)T^d (2p_1)_3\bigg)\bigg]\nonumber\\
\eea

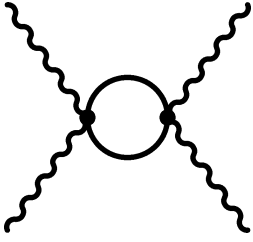
\vspace{0.6mm}
\begin{fmffile}{wboson11}
\begin{fmfgraph}(40,30)
\fmfpen{thick} 
\fmfleft{i1,i2} 
\fmfright{o1,o2}
\fmf{plain,left}{v1,v2}
\fmf{plain,left}{v2,v1}
  \fmf{photon}{i1,v1}
   \fmf{photon}{v2,o1}
   \fmf{photon}{i2,v1}
   \fmf{photon}{v2,o2}
\fmfdotn{v}{2}
  %\fmfdot{v}
  \end{fmfgraph}
\end{fmffile}
\bea
&=&\frac{1}{2}\frac{(2 \pi)^4}{k^4}\int  \frac{d^3p}{(2\pi)^3} \frac{d^3q}{(2\pi)^3} \frac{d^3s}{(2\pi)^3} \frac{d^3p_1}{(2\pi)^3}\frac{1}{p_-q_-s_-(-p-q-s)_-}\frac{1}{(p_1+s)^2}\nonumber\\
&\times&\frac{1}{(p_1+p+q+s)^2}\nonumber\\
&\times& \tr \bigg[\bigg(\cf_{-3}^c(p)\cf_{-3}^d(q)T^cT^d\bigg)\nonumber\\
%&\times&\bigg(\cf_{+-}^b(q) T^b\frac{1}{2}(2p_1+p+q)_-+\cf_{-3}^b(q)T^b(2p_1+p+q)_3\bigg)\nonumber\\
&\times&\bigg(\cf_{-3}^a(s)\cf_{-3}^b(-p-q-s)T^aT^b\bigg)\bigg]\nonumber\\
\eea

\subsection*{Fermionic Theory}
\begin{fmffile}{w6}
\begin{fmfgraph}(40,30)
\fmfpen{thick} 
\fmfleft{i1,i2} 
\fmfright{o1,o2}
\fmf{fermion}{v1,v3}
\fmf{fermion}{v3,v4}
\fmf{fermion}{v4,v2}
\fmf{fermion}{v2,v1}
  \fmf{photon}{i1,v1}
   \fmf{photon}{v3,o1}
   \fmf{photon}{i2,v2}
   \fmf{photon}{v4,o2}
\fmfdotn{v}{4}
  %\fmfdot{v}
  \end{fmfgraph}
\end{fmffile}

\bea
&=&-\frac{1}{4}\frac{(2 \pi)^4}{k^4}\int  \frac{d^3p}{(2\pi)^3} \frac{d^3q}{(2\pi)^3} \frac{d^3s}{(2\pi)^3} \frac{d^3p_1}{(2\pi)^3}\frac{1}{p_-q_-s_-(-p-q-s)_-}\frac{1}{(p_1+p)^2 (p_1+p+q)^2}\nonumber\\
&\times&\frac{1}{(p_1+p+q+s)^2(p_1)^2} \tr \bigg[\bigg(\cf_{+-}^a(p) T^a\gamma_-+\cf_{-3}^a(p)T^a\gamma_3\bigg)\nonumber\\
&\times&\bigg(-i (\gamma_- (p_1+p)_+(1-\lambda^2)+(p_1+p)_-\gamma_++(p_1+p)_3\gamma_3)+\lambda(2 (p_1+p)_+(p_1+p)_-)^{\frac{1}{2}}\bigg)\nonumber\\
&\times&\bigg(\cf_{+-}^b(q) T^b\gamma_-+\cf_{-3}^b(q)T^b\gamma_3\bigg)\nonumber\\
&\times&\bigg(-i (\gamma_- (p_1+p+q)_+(1-\lambda^2)+(p_1+p+q)_-\gamma_++(p_1+p+q)_3\gamma_3)\nonumber\\
&+&\lambda(2 (p_1+p+q)_+(p_1+p+q)_-)^{\frac{1}{2}}\bigg)\nonumber\\
&\times&\bigg(\cf_{+-}^c(s) T^c\gamma_-+\cf_{-3}^c(s)T^c\gamma_3\bigg)\nonumber\\
&\times&\bigg(-i (\gamma_- (p_1+p+q+s)_+(1-\lambda^2)+(p_1+p+q+s)_-\gamma_++(p_1+p+q+s)_3\gamma_3)\nonumber\\
&+&\lambda(2 (p_1+p+q+s)_+(p_1+p+q+s)_-)^{\frac{1}{2}}\bigg)\nonumber\\
&\times&\bigg(\cf_{+-}^d(-p-q-s) T^d \gamma_-+\cf_{-3}^d(-p-q-s)T^d \gamma_3\bigg)\nonumber\\
&\times&\bigg(-i (\gamma_- (p_1)_+(1-\lambda^2)+(p_1)_-\gamma_++(p_1)_3\gamma_3)+\lambda(2 (p_1)_+(p_1)_-)^{\frac{1}{2}}\bigg)
\bigg]\nonumber\\
\eea
\subsection*{Scaling with number of colors N}
$4$-point  amplitudes are also of $\mathcal{O}(\frac{1}{N})$.

\end{appendix}

\bibliographystyle{JHEP}
\bibliography{ONBoso}

\providecommand{\href}[2]{#2}\begingroup\raggedright\begin{thebibliography}{10}

\bibitem{Coleman:1974bu}
S.~R. Coleman, {\it {The Quantum Sine-Gordon Equation as the Massive Thirring
  Model}},  {\em Phys.Rev.} {\bf D11} (1975) 2088.

\bibitem{Mandelstam:1975hb}
S.~Mandelstam, {\it {Soliton Operators for the Quantized Sine-Gordon
  Equation}},  {\em Phys.Rev.} {\bf D11} (1975) 3026.

\bibitem{Witten:1983ar}
E.~Witten, {\it {Nonabelian Bosonization in Two-Dimensions}},  {\em
  Commun.Math.Phys.} {\bf 92} (1984) 455--472.

\bibitem{Burgess:1993np}
C.~Burgess and F.~Quevedo, {\it {Bosonization as duality}},  {\em Nucl.Phys.}
  {\bf B421} (1994) 373--390,
  [\href{http://xxx.lanl.gov/abs/hep-th/9401105}{{\tt hep-th/9401105}}].

\bibitem{Burgess:1994np}
C.~Burgess and F.~Quevedo, {\it {NonAbelian bosonization as duality}},  {\em
  Phys.Lett.} {\bf B329} (1994) 457--462,
  [\href{http://xxx.lanl.gov/abs/hep-th/9403173}{{\tt hep-th/9403173}}].

\bibitem{Burgess:1994tm}
C.~Burgess, C.~Lutken, and F.~Quevedo, {\it {Bosonization in higher
  dimensions}},  {\em Phys.Lett.} {\bf B336} (1994) 18--24,
  [\href{http://xxx.lanl.gov/abs/hep-th/9407078}{{\tt hep-th/9407078}}].

\bibitem{Fradkin:1994tt}
E.~H. Fradkin and F.~A. Schaposnik, {\it {The Fermion - boson mapping in
  three-dimensional quantum field theory}},  {\em Phys.Lett.} {\bf B338} (1994)
  253--258, [\href{http://xxx.lanl.gov/abs/hep-th/9407182}{{\tt
  hep-th/9407182}}].

\bibitem{Giombi:2012ms}
S.~Giombi and X.~Yin, {\it {The Higher Spin/Vector Model Duality}},
  \href{http://xxx.lanl.gov/abs/1208.4036}{{\tt arXiv:1208.4036}}.

\bibitem{Aharony:2011jz}
O.~Aharony, G.~Gur-Ari, and R.~Yacoby, {\it {d=3 Bosonic Vector Models Coupled
  to Chern-Simons Gauge Theories}},  {\em JHEP} {\bf 1203} (2012) 037,
  [\href{http://xxx.lanl.gov/abs/1110.4382}{{\tt arXiv:1110.4382}}].

\bibitem{Aharony:2012nh}
O.~Aharony, G.~Gur-Ari, and R.~Yacoby, {\it {Correlation Functions of Large N
  Chern-Simons-Matter Theories and Bosonization in Three Dimensions}},  {\em
  JHEP} {\bf 1212} (2012) 028, [\href{http://xxx.lanl.gov/abs/1207.4593}{{\tt
  arXiv:1207.4593}}].

\bibitem{Aharony:2012ns}
O.~Aharony, S.~Giombi, G.~Gur-Ari, J.~Maldacena, and R.~Yacoby, {\it {The
  Thermal Free Energy in Large N Chern-Simons-Matter Theories}},
  \href{http://xxx.lanl.gov/abs/1211.4843}{{\tt arXiv:1211.4843}}.

\bibitem{Giombi:2011kc}
S.~Giombi, S.~Minwalla, S.~Prakash, S.~P. Trivedi, S.~R. Wadia, {\em et.~al.},
  {\it {Chern-Simons Theory with Vector Fermion Matter}},  {\em Eur.Phys.J.}
  {\bf C72} (2012) 2112, [\href{http://xxx.lanl.gov/abs/1110.4386}{{\tt
  arXiv:1110.4386}}].

\bibitem{Chang:2012kt}
C.-M. Chang, S.~Minwalla, T.~Sharma, and X.~Yin, {\it {ABJ Triality: from
  Higher Spin Fields to Strings}},
  \href{http://xxx.lanl.gov/abs/1207.4485}{{\tt arXiv:1207.4485}}.

\bibitem{Jain:2012qi}
S.~Jain, S.~P. Trivedi, S.~R. Wadia, and S.~Yokoyama, {\it {Supersymmetric
  Chern-Simons Theories with Vector Matter}},  {\em JHEP} {\bf 1210} (2012)
  194, [\href{http://xxx.lanl.gov/abs/1207.4750}{{\tt arXiv:1207.4750}}].

\bibitem{GurAri:2012is}
G.~Gur-Ari and R.~Yacoby, {\it {Correlators of Large N Fermionic Chern-Simons
  Vector Models}},  {\em JHEP} {\bf 1302} (2013) 150,
  [\href{http://xxx.lanl.gov/abs/1211.1866}{{\tt arXiv:1211.1866}}].

\bibitem{Jain:2013gza}
S.~Jain, S.~Minwalla, and S.~Yokoyama, {\it {Chern Simons duality with a
  fundamental boson and fermion}},  {\em JHEP} {\bf 11} (2013) 037,
  [\href{http://xxx.lanl.gov/abs/1305.7235}{{\tt arXiv:1305.7235}}].

\bibitem{Frishman:2013dvg}
Y.~Frishman and J.~Sonnenschein, {\it {Breaking conformal invariance - Large N
  Chern-Simons theory coupled to massive fundamental fermions}},  {\em JHEP}
  {\bf 1312} (2013) 091, [\href{http://xxx.lanl.gov/abs/1306.6465}{{\tt
  arXiv:1306.6465}}].

\bibitem{Bardeen:2014paa}
W.~A. Bardeen and M.~Moshe, {\it {Spontaneous breaking of scale invariance in a
  D=3 U(N ) model with Chern-Simons gauge fields}},  {\em JHEP} {\bf 1406}
  (2014) 113, [\href{http://xxx.lanl.gov/abs/1402.4196}{{\tt
  arXiv:1402.4196}}].

\bibitem{Bardeen:2014qua}
W.~A. Bardeen, {\it {The Massive Fermion Phase for the U(N) Chern-Simons Gauge
  Theory in D=3 at Large N}},  {\em JHEP} {\bf 1410} (2014) 39,
  [\href{http://xxx.lanl.gov/abs/1404.7477}{{\tt arXiv:1404.7477}}].

\bibitem{Frishman:2014cma}
Y.~Frishman and J.~Sonnenschein, {\it {Large N Chern-Simons with massive
  fundamental fermions - A model with no bound states}},  {\em JHEP} {\bf 12}
  (2014) 165, [\href{http://xxx.lanl.gov/abs/1409.6083}{{\tt
  arXiv:1409.6083}}].

\bibitem{Moshe:2014bja}
M.~Moshe and J.~Zinn-Justin, {\it {3D Field Theories with Chern--Simons Term
  for Large $N$ in the Weyl Gauge}},  {\em JHEP} {\bf 1501} (2015) 054,
  [\href{http://xxx.lanl.gov/abs/1410.0558}{{\tt arXiv:1410.0558}}].

\bibitem{Alvarez:1993qi}
E.~Alvarez, L.~Alvarez-Gaume, J.~Barbon, and Y.~Lozano, {\it {Some global
  aspects of duality in string theory}},  {\em Nucl.Phys.} {\bf B415} (1994)
  71--100, [\href{http://xxx.lanl.gov/abs/hep-th/9309039}{{\tt
  hep-th/9309039}}].

\bibitem{Moreno:2013xya}
E.~Moreno and F.~Schaposnik, {\it {Dualities and bosonization of massless
  fermions in three dimensional space-time}},  {\em Phys.Rev.} {\bf D88}
  (2013), no.~2 025033, [\href{http://xxx.lanl.gov/abs/1303.6488}{{\tt
  arXiv:1303.6488}}].

\bibitem{Redlich:1983kn}
A.~Redlich, {\it {Gauge Noninvariance and Parity Violation of Three-Dimensional
  Fermions}},  {\em Phys.Rev.Lett.} {\bf 52} (1984) 18.

\bibitem{Redlich:1983dv}
A.~Redlich, {\it {Parity Violation and Gauge Noninvariance of the Effective
  Gauge Field Action in Three-Dimensions}},  {\em Phys.Rev.} {\bf D29} (1984)
  2366--2374.

\bibitem{2013PhRvB..87h5132C}
A.~{Chan}, T.~L. {Hughes}, S.~{Ryu}, and E.~{Fradkin}, {\it {Effective field
  theories for topological insulators by functional bosonization}},  {\em PRB}
  {\bf 87} (Feb., 2013) 085132, [\href{http://xxx.lanl.gov/abs/1210.4305}{{\tt
  arXiv:1210.4305}}].

\bibitem{Bralic:1995ip}
N.~Bralic, E.~H. Fradkin, V.~Manias, and F.~A. Schaposnik, {\it {Bosonization
  of three-dimensional nonAbelian fermion field theories}},  {\em Nucl.Phys.}
  {\bf B446} (1995) 144--158,
  [\href{http://xxx.lanl.gov/abs/hep-th/9502066}{{\tt hep-th/9502066}}].

\bibitem{Klebanov:2002ja}
I.~Klebanov and A.~Polyakov, {\it {AdS dual of the critical O(N) vector
  model}},  {\em Phys.Lett.} {\bf B550} (2002) 213--219,
  [\href{http://xxx.lanl.gov/abs/hep-th/0210114}{{\tt hep-th/0210114}}].

\bibitem{Sezgin:2002rt}
E.~Sezgin and P.~Sundell, {\it {Massless higher spins and holography}},  {\em
  Nucl.Phys.} {\bf B644} (2002) 303--370,
  [\href{http://xxx.lanl.gov/abs/hep-th/0205131}{{\tt hep-th/0205131}}].

\bibitem{Dunne:1998qy}
G.~V. Dunne, {\it {Aspects of Chern-Simons theory}},
  \href{http://xxx.lanl.gov/abs/hep-th/9902115}{{\tt hep-th/9902115}}.

\bibitem{Koch:2010cy}
R.~d.~M. Koch, A.~Jevicki, K.~Jin, and J.~P. Rodrigues, {\it {$AdS_4/CFT_3$
  Construction from Collective Fields}},  {\em Phys.Rev.} {\bf D83} (2011)
  025006, [\href{http://xxx.lanl.gov/abs/1008.0633}{{\tt arXiv:1008.0633}}].

\bibitem{Rocek:1991ps}
M.~Rocek and E.~P. Verlinde, {\it {Duality, quotients, and currents}},  {\em
  Nucl. Phys.} {\bf B373} (1992) 630--646,
  [\href{http://xxx.lanl.gov/abs/hep-th/9110053}{{\tt hep-th/9110053}}].

\bibitem{Nair:2013hpa}
V.~P. Nair, {\it {The Hamiltonian Approach to Yang-Mills (2+1): An Update and
  Corrections to String Tension}},  {\em J. Phys. Conf. Ser.} {\bf 462} (2013),
  no.~1 012039.

\bibitem{delaOssa:1992vc}
X.~C. de~la Ossa and F.~Quevedo, {\it {Duality symmetries from nonAbelian
  isometries in string theory}},  {\em Nucl. Phys.} {\bf B403} (1993) 377--394,
  [\href{http://xxx.lanl.gov/abs/hep-th/9210021}{{\tt hep-th/9210021}}].

\bibitem{Lozano:2011kb}
Y.~Lozano, E.~O~Colgain, K.~Sfetsos, and D.~C. Thompson, {\it {Non-abelian
  T-duality, Ramond Fields and Coset Geometries}},  {\em JHEP} {\bf 06} (2011)
  106, [\href{http://xxx.lanl.gov/abs/1104.5196}{{\tt arXiv:1104.5196}}].

\end{thebibliography}\endgroup

\end{document}